\documentclass[11pt]{article}
\usepackage{graphicx,amsmath,amssymb,epsf, hyperref} 
\usepackage{latexsym,bm,slashed} 
\usepackage{xcolor} \definecolor{dark}{rgb}{0.10,0.2,0.3}
\definecolor{magenta}{rgb}{0.7,0.1,0.3}
\definecolor{purpure}{rgb}{0.5,0.15,0.3}
\usepackage[font=small,format=plain,labelfont=bf,up,textfont=it,up]{caption}
\usepackage{hyperref, cite} \hypersetup{colorlinks, citecolor=blue,
  filecolor=blue, linkcolor=magenta,
  urlcolor=purpure,hyperfootnotes=true,pdftex}

 \title{\bf \Large Ratio of $J/\Psi$ and $\Psi(2s)$ exclusive photoproduction cross-sections as an indicator for the presence of non-linear QCD evolution}
 \author{Marco Alcazar Peredo and  Martin Hentschinski\\ \\ 
Departamento de Actuaria, F\'isica y Matem\'aticas,
Universidad de las Americas Puebla, \\ Santa Catarina Martir, 72820 Puebla, Mexico }

\begin{document}

\maketitle
\begin{abstract}
 We investigate the proposal that the rise with energy of the ratio of the exclusive photo-production cross-sections of vector mesons $\Psi(2s)$ and $J/\Psi$ can serve as an indicator for the presence of high gluon densities and associated non-linear high energy evolution; we  study this proposal  for both photoproduction on  a  proton and a lead nucleus. While previous studies were based on unintegrated gluon distributions subject to linear (Balitsky-Fadin-Kuraev-Lipatov) and non-linear (Balitsky-Kovchegov) evolution equations, the current study is based on the Golec-Biernat W\"usthoff (GBW) and Bartels Golec-Biernat Kowalski (BGK) models, which allow assessing more directly  the relevance of non-linear corrections for the description of the energy dependence of the photoproduction cross-section. We find that the rise of the ratio is directly related to the presence of a node in the $\Psi(2s)$ wave function and only manifests itself for the complete non-linear models, while it is absent for their linearized versions. We further provide predictions based on leading order collinear factorization and examine to which extent such an approach can mimic a ratio rising with energy. We also provide a description of recent ALICE data on the energy dependence of the photonuclear $J/\Psi$ production cross-section and give predictions for the energy dependence of the ratio of $\Psi(2s)$  and $J/\Psi$ photoproduction cross-sections for both scattering on a proton and a lead nucleus.
 \end{abstract}

\section{Introduction}
\label{sec:intro}

Exclusive photo-production of charmonium in ultra-peripheral
collisions at the Large Hadron Collider (LHC) provides one of the few
processes at the LHC that allows for an exploration of Quantum
Chromodynamics (QCD) at ultra-small values of $x$, i.e., the
perturbative high energy limit of strong interactions. Here,
$x = M_V^2/W^2$, where $M_V$ denotes the mass of the vector meson and
$W$ is the center of mass energy of the photon-proton or
photon-nucleus reaction.  While a power-like rise of the gluon
distribution with decreasing $x$ has been both observed at the HERA
collider and is theoretically predicted by
Balitsky-Fadin-Kuraev-Lipatov (BFKL) evolution
\cite{Kuraev:1977fs,Kuraev:1976ge, Balitsky:1978ic}, it is known that
such a rise cannot continue down to arbitrarily small values of
$x$. At some value of $x$, this power-like rise must slow down and
eventually saturate \cite{Gribov:1984tu, Gelis:2010nm,
  Morreale:2021pnn}. While there exist strong theoretical arguments for
saturation of gluon densities \cite{Balitsky:1995ub,
  Jalilian-Marian:1997ubg, Kovchegov:1999yj, Iancu:2000hn,
  Weigert:2000gi, Iancu:2001ad, Ferreiro:2001qy}, it is still needed
to pin down the region where such dynamics turns relevant for the
description of collider data and to provide clear evidence for the
emergence of non-linear QCD evolution in data.

Reactions which involve charmed final states are for such explorations of special interest, see e.g.,  \cite{Celiberto:2023fzz,  Chapon:2020heu} for inclusive examples, since  the charm mass provides a hard scale of the order of the charm mass $m_c \simeq 1.4$~GeV which places the reaction on the boundary between perturbative and non-perturbative QCD dynamics. A class of reactions which has drawn particular interest is in this context photon induced production of charmed vector mesons in exclusive reactions, see e.g. \cite{Cepila:2018faq, Krelina:2019gee,  Klein:2019qfb, Kopeliovich:2020has,  Bendova:2020hbb, Jenkovszky:2021sis, Flett:2021xsl, Mantysaari:2021ryb, Mantysaari:2022kdm,  Goncalves:2022ret, Wang:2022vhr} for recent theory and \cite{Klein:2020nvu, Bylinkin:2022wkm, ALICE:2023fov} for  experimental proposals; see also the reviews \cite{Amoroso:2022eow, Hentschinski:2022xnd, Frankfurt:2022jns}. Once  HERA and LHC data are combined, such reactions allow to explore low $x$ dynamics from intermediate values $x\simeq 10^{-2}$  down to  $x =10^{-6}$. The low $x$ description is therefore probed over several orders in magnitude.

In  \cite{Hentschinski:2020yfm}, see also \cite{Garcia:2019tne, Bautista:2016xnp}, the energy dependence of the photoproduction cross-section has been studied through comparing fits of unintegrated gluon distributions subject to   next-to-leading order (NLO) BFKL evolution \cite{Hentschinski:2012kr,Hentschinski:2013id, Chachamis:2015ona} (Hentschinski Salas Sabio-Vera; HSS) and DGLAP improved BK evolution \cite{Kutak:2012rf} (Kutak Sapeta; KS) for the case of exclusive photoproduction on a proton, see also \cite{Cepila:2020uxc} for a related study. In this way one compares directly linear and non-linear low $x$ frameworks with the possibility to distinguish between both descriptions in a comparison to data. While central values of results based on linear and non-linear unintegrated gluon distributions  begin to differ for $x <10^{-4}- 10^{-5}$, associated uncertainties do not allow to draw any substantial conclusions on the presence/absence of non-linear QCD dynamics. It was however realized that the ratio of $\Psi(2s)$ and $J/\Psi$ production cross-section yields a characteristically different energy dependence for linear (BFKL) and non-linear (BK) evolution equations: the ratio rises with energy for full non-linear QCD evolution while it is approximately constant for linear evolution. 

While the rise of the ratio of $\Psi(2s)$ and $J/\Psi$ photoproduction cross-sections with energy has been noted before,  if the energy dependence is described through dipole cross-sections which eventually unitarize, see e.g. \cite{Nemchik:1997xb, Cepila:2019skb}, the observation that linear QCD evolution is associated with a constant ratio was not studied so far, to the best of our knowledge; similar ideas have been however used in \cite{STAR:2006dgg} in the context of   inclusive hadron production in $pA$  or  \cite{Dominguez:2011cy}  in $J/\Psi$ gluoproduction to test  gluon saturation models. 
While the result found for   $\Psi(2s)$ and $J/\Psi$ photoproduction is at first encouraging  and would imply certain  benefits  for the search for signals of non-linear QCD dynamics at current and future collider experiments, see e.g. \cite{Bylinkin:2022wkm}, it is at the same time  not clear whether the observed behavior can be directly associated with the presence/absence of non-linear QCD dynamics or whether it is rather an artefact of the KS and HSS unintegrated gluon distributions. To clarify this point further, we  provide in this paper a study of the same phenomena using (semi-)analytical dipole models, which allow for a more straightforward manipulation of linear vs. non-linear effects and provide in this way  a more direct study of their impact on  observables. At the same time it must be however noted that dipole models only capture certain features of the solution of non-linear QCD evolution equations. There is therefore no guarantee that they provide a correct prediction, in particular if their region of applicability is extended beyond the region of phase space  for which their parameters have been originally fitted to\footnote{A similar statement applies if dipole models fitted to proton data are extended to nuclear targets}. Their advantage is that they provide direct analytical access to the underlying dynamics which generates non-linear QCD evolution and therefore allows to test in a more direct way the relevance of such effects and underlying concepts.

We therefore repeat in this paper the program of \cite{Hentschinski:2020yfm}, while we model the gluon density through  two dipole models, the Golec-Biernat W\"usthoff (GBW) model \cite{Golec-Biernat:1998zce} and Bartels Golec-Biernat Kowalski (BGK) model \cite{Bartels:2002cj}, which we treat both in their linearized and complete exponentiated, i.e. unitarized, versions, see below for the precise definitions. We further introduce a new parameter into these models, which allows us to vary the strength of potential non-linear effects. While our study focuses on the study of the ratio of $\Psi(2s)$ and $J/\Psi$ photo-productions in photon-proton collisions, we also give a description of the very recent ALICE  and CMS data on $J/\Psi$ photo-productions in photon-lead collisions \cite{ALICE:2023jgu,CMS:2023snh} based on these models and provide predictions for $\Psi(2s)$ photonuclear production as well as  for the ratio $\Psi(2s)$ over $J/\Psi$.

The outline of this paper is as follows:  In Sec.~\ref{sec:setup}, we give an overview of the theoretical setup of our study and present numerical results for the energy dependence of photoproduction cross-sections. Sec.~\ref{sec:observable} identifies the geometric scaling region as the relevant region for the detection of non-linear dynamics and explores to which extent the ratio of  $\Psi(2s)$ and $J/\Psi$ photoproduction cross-sections is able to capture this type of dynamics; we further present here our predictions for the cross-section ratio. We draw our conclusions in Sec.~\ref{sec:concl}.

\section{Photoproduction cross-sections}
\label{sec:setup}

\begin{figure}[t]
  \centering
   \includegraphics[width = .5\textwidth]{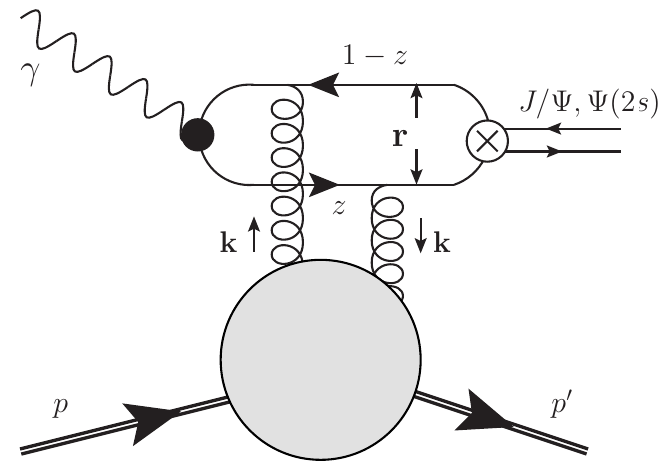}
  \caption{Exclusive photoproduction of vector mesons $J/\Psi$ and
    $\Psi(2s)$. For the quark-anti quark dipole, we indicate photon momentum
    fractions $z$ and $1-z$ as well as the transverse separation ${\bm
    r}$. Finally ${\bm k}$ denotes the transverse momentum transmitted
from the unintegrated gluon distribution of the proton; the latter is indicated
through the gray blob.}
  \label{fig:reaction}
\end{figure}

We  study the process, 
\begin{align}
  \label{eq:30}
 \gamma(q) + p(p) & \to V(q') + p(p'), & V &= J/\Psi, \Psi(2S) \, ,
\end{align}
where  $\gamma$ denotes a quasi-real
photon with virtuality $Q \simeq 0$, which stems from an electron (HERA) or a proton/lead nucleus in the case of LHC data; $W^2 = (q + p)^2$ is the squared
center-of-mass energy of the $\gamma(q) + p(p)$ reaction, see also Fig.~\ref{fig:reaction}. With  $t = (q - q')^2$, the differential cross-section for
the exclusive photo-production of a vector meson can be written as
\begin{align}
  \label{eq:16}
  \frac{d \sigma}{d t} \left(\gamma p \to V p \right) 
& = 
\frac{1}{16 \pi} \left|\mathcal{A}_{T}^{\gamma p \to V p}(x, t) \right|^2  , & V &= J/\Psi, \Psi(2S) \, .
\end{align}
Here $\mathcal{A}_T(W^2, t)$ denotes the scattering amplitude for a transverse polarized real photon with color singlet exchange in the
$t$-channel, with an overall factor $W^2$ already extracted.  $x = M_V^2/W^2$ with $M_V$ the mass of the vector meson; see also  {\it e.g.} \cite{Bautista:2016xnp}.  To access the inclusive gluon distribution, we 
follow a two-step procedure, frequently employed in the
literature, see e.g. \cite{Kowalski:2006hc, Marquet:2007nf, Jones:2013pga}:  First  one determines the differential cross-section at zero
momentum transfer $t=0$, which can be related to  the
inclusive gluon distribution; in  a second step  one uses to ansatz  $d\sigma/dt \sim  \exp\left[-|t| B_D\right]$ with the diffractive slope parameter $B_D$, which can be extracted from the  $t$-dependence measured at HERA and LHC experiments. Given this dependence, the total cross-section   is obtained as 
\begin{align}
  \label{eq:16total}
 \sigma^{\gamma p \to V p}(W^2) & = \frac{1}{B_D(W)} \frac{d \sigma}{d t} \left(\gamma p \to V p \right)\bigg|_{t=0}
.
\end{align}
The  uncertainty introduced through the modeling of
the $t$-dependence mainly affects the
overall normalization of the cross-section with a possible mild logarithmic
dependence on the energy.
For photon-proton collisions, we employ  an energy-dependent
diffractive slope  $B_D$,
\begin{align}
  \label{eq:18}
  B_{D,p}(W) & =\left[  b_{0,p} + 4 \alpha'_{p} \ln \frac{W}{W_0} \right] \text{GeV}^{-2}.
\end{align}
where  $W_0 = 90$~GeV and  $b_{0,p}$ and $\alpha'_p$ determined in \cite{Cepila:2019skb} from a fit to HERA data with $b_{0,p}^{J/\Psi} = 4.62$, $b_{0,p}^{\Psi(2s)} = 4.86$, while $\alpha_p^{'J/\Psi} = 0.171$ and  $\alpha_p^{'\Psi(2s)} = 0.151$. 
To  compare to recent ALICE data for photonuclear $J/\Psi$ production \cite{ALICE:2023jgu}, we need to determine the diffractive slope for this reaction. Since the currently available data set is integrated over the rapidity of the $J/\Psi$ vector meson and therefore exhibits no $W$-dependence, we chose the ansatz
\begin{align}
  \label{eq:slope}
  \frac{d\sigma}{d|t|} & = C \cdot e^{- B_D |t|},
\end{align}
and find $C = (1.62\pm 0.08 )$b$/$GeV$^2$ and $B_D =  (4.01\pm 0.15 )\cdot 10^{2}/$GeV$^2$  with $\chi^2/$d.o.f$=0.97$ from a fit to  the first measurement of the $|t|$ dependence of the coherent $J/\Psi$ photonuclear production, obtained by the ALICE collaboration in  \cite{ALICE:2021tyx},
\begin{figure}[t]
  \centering
  \includegraphics[width=.65\textwidth]{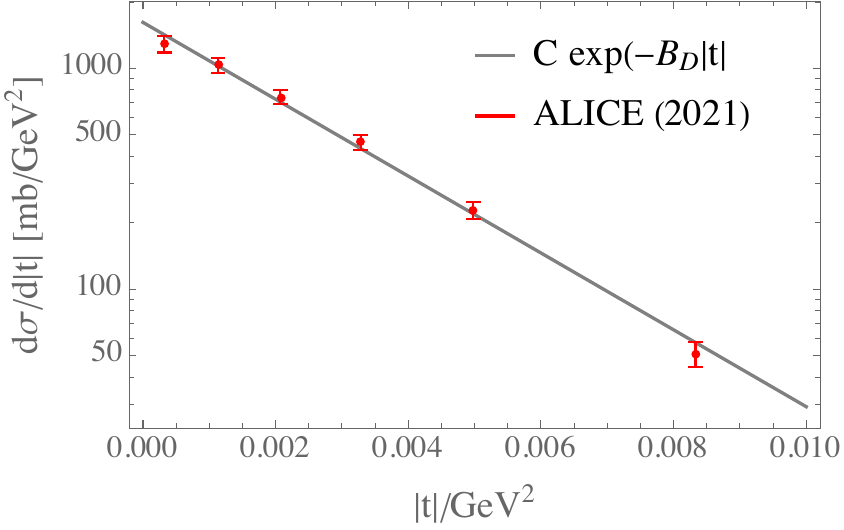}
 
  \caption{Diffractive slope for photonuclear reactions as obtained from a fit to Eq.~\eqref{eq:slope}. We also show ALICE data  \cite{ALICE:2021tyx}. }
  \label{fig:slopeNUCj}
\end{figure}
see also Fig.~\ref{fig:slopeNUCj} for a comparison of data to the employed parametrization; when preparing predictions for the $\Psi(2s)$ photonuclear production cross-section, we will also make use of this result. 

Apart from a potential mild energy dependence of the slope parameter, the entire energy dependence is then contained in the scattering amplitude. 
Its  dominant contribution is provided by the imaginary
part, while corrections due to the real part of the scattering amplitude
can be estimated using dispersion relations, 
\begin{align}
  \label{eq:32}
  \frac{\Re\text{e} \mathcal{A}(x)}{\Im\text{m} \mathcal{A}(x)}
&=
\tan \frac{\lambda(x) \pi }{2},
&
 \lambda(x) & = 
\frac{d \ln \Im\text{m}  \mathcal{A}(x) }{ d \ln 1/x} \, .
\end{align}
As noted in \cite{Bautista:2016xnp}, the dependence of the slope
parameter $\lambda$ on $x$ provides a sizable correction to
the  $W$ dependence of the complete
cross-section. We therefore do not assume $\lambda =$const., but instead determine the slope $\lambda$ directly from the $x$-dependent imaginary part of the scattering amplitude.

\subsection{Wave function overlap}
\label{sec:WFO}

Within high energy factorization, the imaginary part of the scattering amplitude at $t=0$ is obtained as a convolution of the light-front wave function -- which describes the formation of a color dipole and its subsequent transition into a vector meson -- and the dipole cross-section.   
In the following, we use  a simple Gaussian model for the vector meson  wave function, which yields
\begin{align}
\label{am-i}  
\Im \text{m}\mathcal{A}^{\gamma p\rightarrow
    Vp}_{T} (x, t=0) &= \int\!d^2{\bm r}
  \Sigma(r) \,
  \sigma_{q\bar{q}}\left(x,r\right)\,, 
\end{align}
where $r = {\bm r}$ denotes the transverse separation of the dipole and 
\begin{align}
  \label{eq:21}
\Sigma(r) &=  \int_0^1\! \frac{d {z}}{4 \pi }\; \left(\Psi_{V}^{*}\Psi_T\right)(r,z) \notag \\
&  =  \int_0^1\! \frac{d {z}}{4 \pi }\; \frac{\hat{e}_f eN_c}{\pi z (1-z)}
 \bigg\{
 m_f^2 K_0(m_c r) \phi_T(r,z) - \left[z^2 + (1-z)^2 \right] \epsilon K_1(m_c r) \partial_r \phi_T(r,z)
\bigg\}
 \, .
\end{align}
 Note that \cite{Hentschinski:2020yfm} uses a refined description for the wave function overlap based on the setup and wave functions presented in \cite{Krelina:2018hmt,Cepila:2019skb}; as an alternative to these wave functions one might also consider the set presented in \cite{Li:2017mlw}. We found that the effects of this improved description are small for the current study. We  therefore stick for simplicity with the simple boosted  Gaussian model according to the  Brodsky-Huang-Lepage prescription
\cite{Brodsky:1980vj, Cox:2009ag, Nemchik:1994fp},
\begin{align}
\label{eq:1s2s_groundstate}
\phi_{T}^{1s}(r,z) &= \mathcal{N}_{T, 1s} z(1-z)
  \exp\left(-\frac{m_f^2 \mathcal{R}_{1s}^2}{8z(1-z)} - \frac{2z(1-z)r^2}{\mathcal{R}_{1s}^2} + \frac{m_f^2\mathcal{R}_{1s}^2}{2}\right)  \, ,
\\
  \label{eq:2s_excitedstate}
  \phi_{T,L}^{2s}(r,z) &= \mathcal{N}_{T, 2s} z(1-z)
  \exp\left(-\frac{m_f^2 \mathcal{R}_{2s}^2}{8z(1-z)} - \frac{2z(1-z)r^2}{\mathcal{R}_{2s}^2} + \frac{m_f^2\mathcal{R}_{2s}^2}{2}\right) \notag \\
& \cdot \left[ 1 + \alpha_{2s} \left(2 + \frac{m_f^2 \mathcal{R}_{2s}^2}{4 z(1-z)} - \frac{4 z(1-z) r^2}{\mathcal{R}_{2s}^2} - m_f^2 \mathcal{R}_{2s}^2\right)\right]\,.
\end{align}
The free parameters of this parametrization have been
determined in various studies from the  normalization and orthogonality of the wave functions as well as  the decay width of the vector mesons. In the
following we use the values found in 
\cite{Armesto:2014sma} which we summarize in Tab.~\ref{vm_fit}.
\begin{table}{t}
\centering
\caption{Parameters of the boosted Gaussian vector meson wave functions for $J/\Psi$ and $\Psi(2s)$  \cite{Armesto:2014sma}.}

\begin{tabular}{c|c|c|c|c|c}
\hline\hline &&&& \vspace{-.2cm}\\
Meson & $m_f/\text{GeV}$  & $\mathcal{N}_T$ &  $\mathcal{R}^2$/$\text{GeV}^{-2}$
 & $\alpha_{2s}$ & $M_V$/GeV   
\\
&&&&& \vspace{-.2cm}\\  \hline
$J/\Psi$ & $m_c=1.4$&   $0.596$ & $2.45$ & $-$  &$3.097$      \\ \hline
$\Psi(2s)$ & $m_c = 1.4 $ & $0.67$ & $3.72$ &$ -0.61$ & $3.686$   \\ \hline
\end{tabular}
\label{vm_fit}
\end{table}

\subsection{Dipole cross-sections}
\label{sec:gluon}

Within collinear factorization, one finds to leading order for the dipole cross-section \cite{Frankfurt:1996ri, Bartels:2002cj}
\begin{align}
  \label{eq:sigcoll}
   \sigma_{q\bar{q}}^{\text{collinear}}(x, r) & = \frac{\pi^2}{3} r^2 \alpha_s(\mu^2) xg(x, \mu^2).
\end{align}
The renormalization scale $\mu$ is usually identified with the factorization scale and taken to depend on the dipole size with $\mu^2 \sim 1/r^2$ for small dipole sizes; $xg(x, \mu^2)$ denotes the collinear gluon distribution subject to leading order DGLAP evolution. A simple way to estimate corrections that yield  unitarization of this dipole cross-section in the limit of large dipole separations $r$ and/or large gluon densities is to  exponentiate the collinear cross-section, which yields the   Bartels--Golec-Biernat--Kowalski (BGK) model,
\begin{align}
  \label{eq:sigBGK}
   \sigma_{q\bar{q}}^{\text{BGK}}(x, r) & = \sigma_0^{\text{BGK}} \left[ 1 -  \exp \left(-\frac{r^2 \pi^2\alpha_s(\mu_r^2) xg(x, \mu_r^2) }{3 \sigma_0^{\text{BGK}} }  \right)\right].
\end{align}
 See \cite{Frankfurt:2022jns, Goda:2022wsc} for a recent comprehensive review. The above exponentiation introduces a new parameter, $\sigma_0$, which yields the value of the dipole cross-section in the black disk limit, corresponding to the transverse size of the target. An even simpler model is provided by the  Golec-Biernat, W\"usthoff (GBW) model,
\begin{align}
  \label{eq:sigGBW}
  \sigma_{q\bar{q}}^{\text{GBW}}(x, r) & = \sigma_0^{\text{GBW}} \left[1- \exp \left(-\frac{r^2 Q_s^2(x)}{4}  \right)\right], & 
 Q_s^2(x) & = Q_0^2\left(\frac{x_0}{x}\right)^\lambda,
\end{align}
where $Q_s$ denotes the saturation scale within the model and gathers various factors of the collinear cross-section into a single factor. Both models have been recently refitted for dipole scattering on a proton  to combined HERA data in \cite{Golec-Biernat:2017lfv} where free parameters are obtained as  $\sigma_0^{\text{GBW}} = (27.43 \pm 0.35)$~mb,  $\lambda = 0.248\pm 0.002$,  $x_0 = (0.40 \pm 0.04)\cdot 10^{-4}$, while $Q_0 = 1$~GeV for the GBW model. For the BGK model,  $g(x, \mu^2)$ is  subject to leading order DGLAP evolution equation without quarks,
\begin{align}
  \label{eq:dglap}
  \frac{d}{d\mu^2} g(x, \mu^2) & = \frac{\alpha_s}{2 \pi} \int_x^1\frac{dz}{z} P_{gg}(z) g(x/z, \mu^2), & xg(x, Q_0^2) & = A_g x^{-\lambda_g} (1-x)^{5.6},
\end{align}
where $xg(x, Q_0^2)$ denotes the gluon distribution at the initial scale $Q_0 = 1$~GeV. Following the recent fit \cite{Golec-Biernat:2017lfv} of this model, we evaluate the  gluon distribution  and the QCD running coupling at the dipole size-dependent scale
\begin{align}
  \label{eq:muR}
  \mu_r^2 & = \frac{\mu_0^2}{1 - \exp\left(- \mu_0^2 r^2/C \right)}.
\end{align}
The remaining parameters of the model have been obtained as    $\sigma_0^{\text{BGK}} = (22.93 \pm 0.27)$~mb,  $A_g = 1.07 \pm 0.13$,  $\lambda_g = 0.11 \pm 0.03$,  $C = 0.27 \pm 0.04$,  $\mu_0^2 = (1.74 \pm 0.16 )$~GeV$^2$. The exponentiated terms allow us within these simple models to explore the relevance of non-linear QCD dynamics for the description of data. We will therefore refer in the following to these exponentiated terms also as the `non-linear' corrections.\\

\subsection{Modified Dipole Cross-sections}
\label{sec:mod}

To explore the relevance of the exponentiated terms, which simulate non-linear QCD evolution, we will compare for the following numerical study  both complete and linearized models. In addition, we introduce in the following a parameter `$k$'  which allows for a smooth transition between both scenarios, i.e., which allows to vary the `density' of gluons by hand. We  introduce this parameter $k$ through a rescaling $Q_s^2(x) \to k \cdot Q_s^2(x)$, while we keep the linearized dipole cross-sections fixed. For the GBW model this leads to 
 \begin{align}
  \label{eq:sigGBWhk}
  \sigma_{q\bar{q}}^{\text{GBW}}(x,r,k) & =
 \sigma_0^{\text{GBW}}  Q_s^2(x) \left(\frac{r^2}{4}\right) \left[1 + \sum_{n=1}^\infty \frac{1}{(n+1)!} \left(-k \cdot \frac{r^2 Q_s^2(x)}{4}\right)^n \right]
\notag \\
&=
 \frac{\sigma_0^{\text{GBW}}}{k}  \left[1 - \exp\left(-k\cdot \frac{r^2 Q_s^2(x)}{4}\right) \right] .
\end{align}
With this modification, $k=0$ corresponds to the linear case, whereas $k=1$  yields the current HERA fit of the model; finally, $k> 1$ implies an additional enhancement of non-linear effects. Within this simple approach, $k$ can be understood as a parameter that controls the strength of the triple Pomeron vertex and, therefore, the relevance of non-linear dynamics. We also apply an identical modification to  the BGK model, 
\begin{align}
  \label{eq:sigBGKk}
   \sigma_{q\bar{q}}^{\text{BGK}}(x, r, k) & = \frac{\sigma_0^{\text{BGK}}}{k} \left[ 1 -  \exp \left(-k\cdot \frac{ r^2 \pi^2\alpha_s(\mu_r^2) xg(x, \mu_r^2) }{3 \sigma_0^{\text{BGK}} }  \right)\right].
\end{align}

\subsection{Nuclear effects}
\label{sec:photonnuclear}

If the color dipole scatters on a large nucleus instead of a single proton,  one expects an increase of the saturation scale due to the   nuclear ``oomph factor'',
\begin{align} 
 \label{eq:oomph}
  Q_{s, A}^2(x) & \simeq A^{\frac{1}{3}} Q_{s}^2(x),
\end{align}
where $A$ denotes the number of nucleons in the nucleus, and $Q_{s,A}^2(x)$ is the saturation scale for the nuclear target, while $Q_{s}^2(x)$ denotes the saturation scale for a single proton, as obtained from the fit to HERA data.   With the transverse size of the dipole cross-section  scaling as $\sim A^{2/3}$, we finally obtain
\begin{align}
  \label{eq:qqb_A}
  \sigma_{q\bar{q}; A}(x, r) & = A  \sigma_{q\bar{q}}(x, r, k= A^{\frac{1}{3}}),
\end{align}
where we use the $k$ parameter of  Eqs.~\eqref{eq:sigGBWhk}, \eqref{eq:sigBGKk} to implement the nuclear enhancement of the saturation scale, corresponding to an increase in the density of gluons. Concluding we would like to stress that the above way to introduce nuclear dependence is of course a very crude approach. It neither includes possible variations of the saturation scale at different impact parameters nor does it include the evolution of the latter with energy. Such effects have been already studied in the literature, see for instance   \cite{Azevedo:2022ozu, Mantysaari:2018nng}  for studies at the level of dipole models or \cite{Cepila:2018faq,Cepila:2020uxc,Bendova:2020hbb} who study  $J/\Psi$ photoproduction on a lead target using  a solution to impact parameter dependent BK evolution, with initial conditions fitted to nuclear PDFs. Such an approach is clearly technically more advanced than the above treatment and in this sense superior to our results. Absence of such effects in our model might therefore in principle spoil our results. Our approach has on the other the advantage that it provides an extremely simply way to test the basic ideas of gluon saturation, such as  an $A^{\frac{1}{3}}$ enhancement of color sources in the large nucleus,  and therefore allows for a simple test  of the basic ideas underlying gluon saturation.

\subsection{Results for the photo-production cross-section}
\label{sec:Xsec}

\begin{figure}[t]
  \centering
  \parbox{.49\textwidth}{
     \includegraphics[width=.45\textwidth]{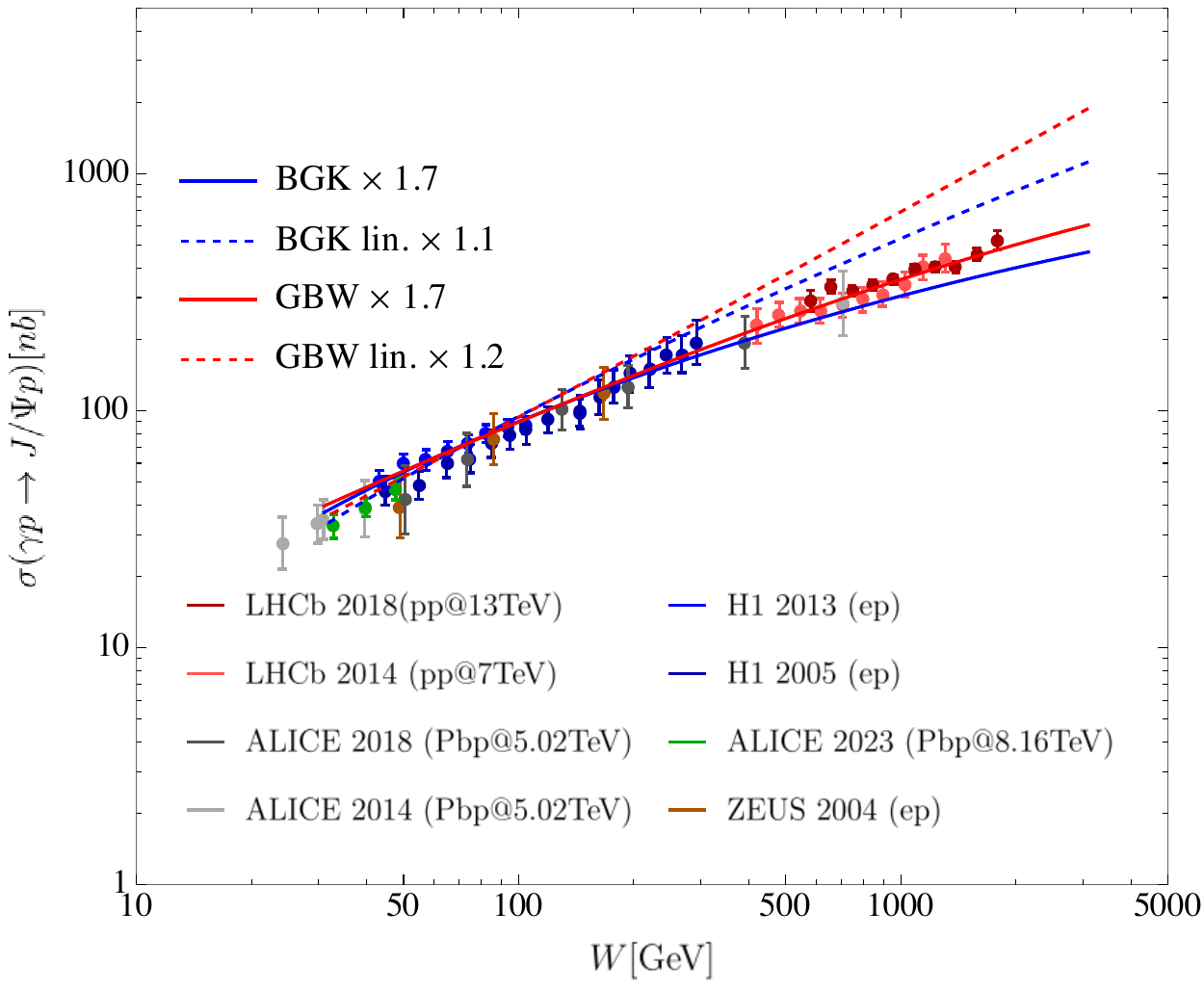}
     }
 \parbox{.49\textwidth}{
     \includegraphics[width=.45\textwidth]{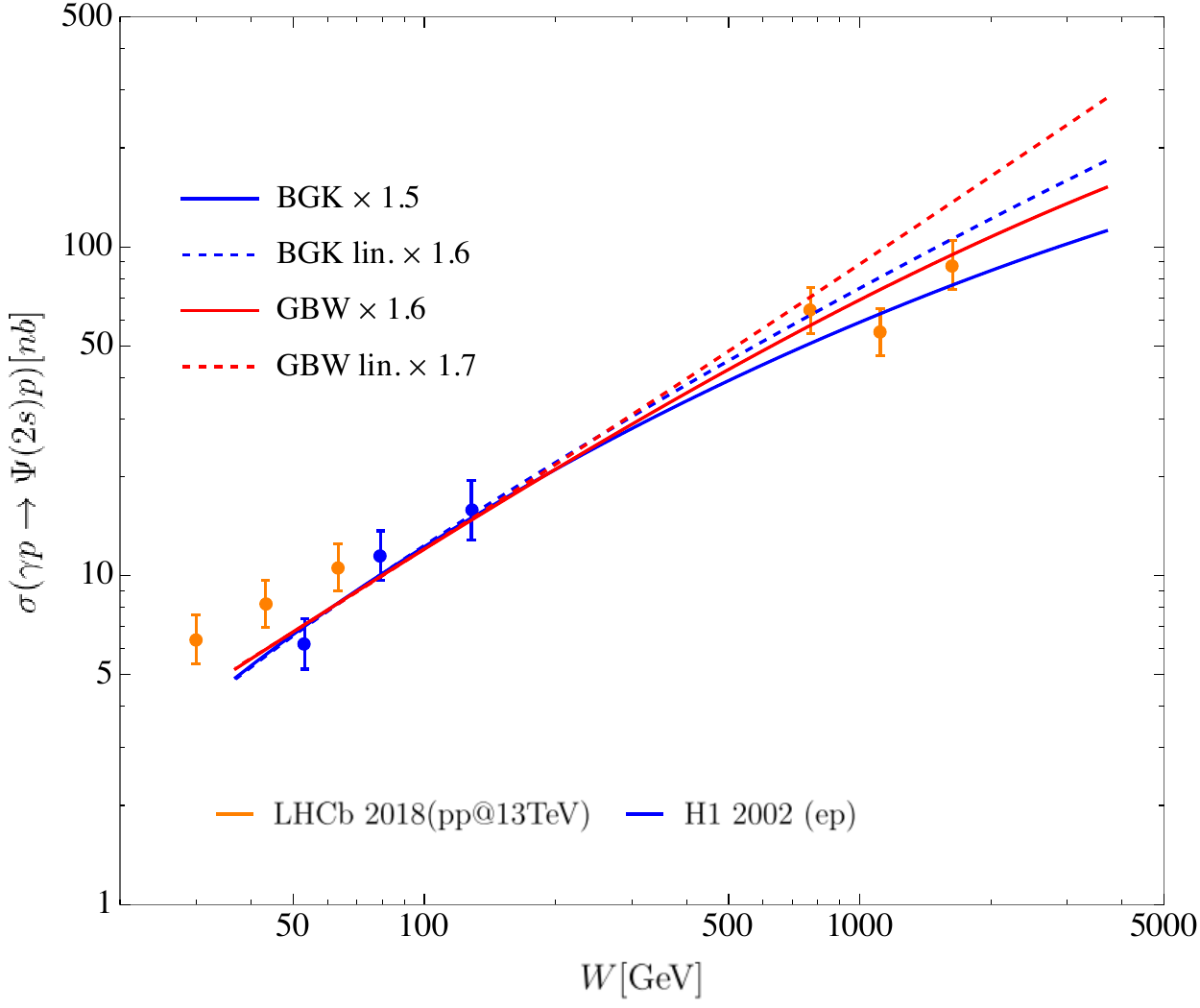}
     }
  \caption{Energy dependence of the total cross-sections for exclusive photoproduction of $J/\Psi$ (left) and $\Psi(2s)$(right) as obtained within the dipole models discussed in the text.  We further display photo-production data measured at HERA by ZEUS
    \cite{Chekanov:2004mw} and H1
    \cite{Alexa:2013xxa,Aktas:2005xu} as well as LHC data obtained
    from ALICE \cite{TheALICE:2014dwa, Acharya:2018jua, ALICE:2023mfc} and LHCb
     \cite{Aaij:2013jxj, Aaij:2018arx} for $J/\Psi$ production as well as H1
    \cite{Schmidt:2001tu, Adloff:2002re} and   LHCb data
    \cite{ Aaij:2018arx} for the $\Psi(2s)$ photoproduction cross-section.}
  \label{fig:full_vs_linear}
\end{figure}
In Fig.~\ref{fig:full_vs_linear} we compare the energy dependence of the complete dipole models against their corresponding linearized version, with the overall normalization fitted to low energy HERA data. While $J/\Psi$ tends to prefer the complete saturation models which include non-linear effects, the uncertainties of the LHCb data shown in the comparison are likely to be underestimated, since they only include the experimental uncertainties of the distribution in rapidity, but not effects due to different methods of unfolding the photon-proton cross-section from  proton-proton data. Moreover, both models have been fitted in their complete non-linear version to HERA data and it is therefore natural for the non-linear version to align better with data. 
\begin{figure}[t]
  \centering
  \includegraphics[width=.49\textwidth]{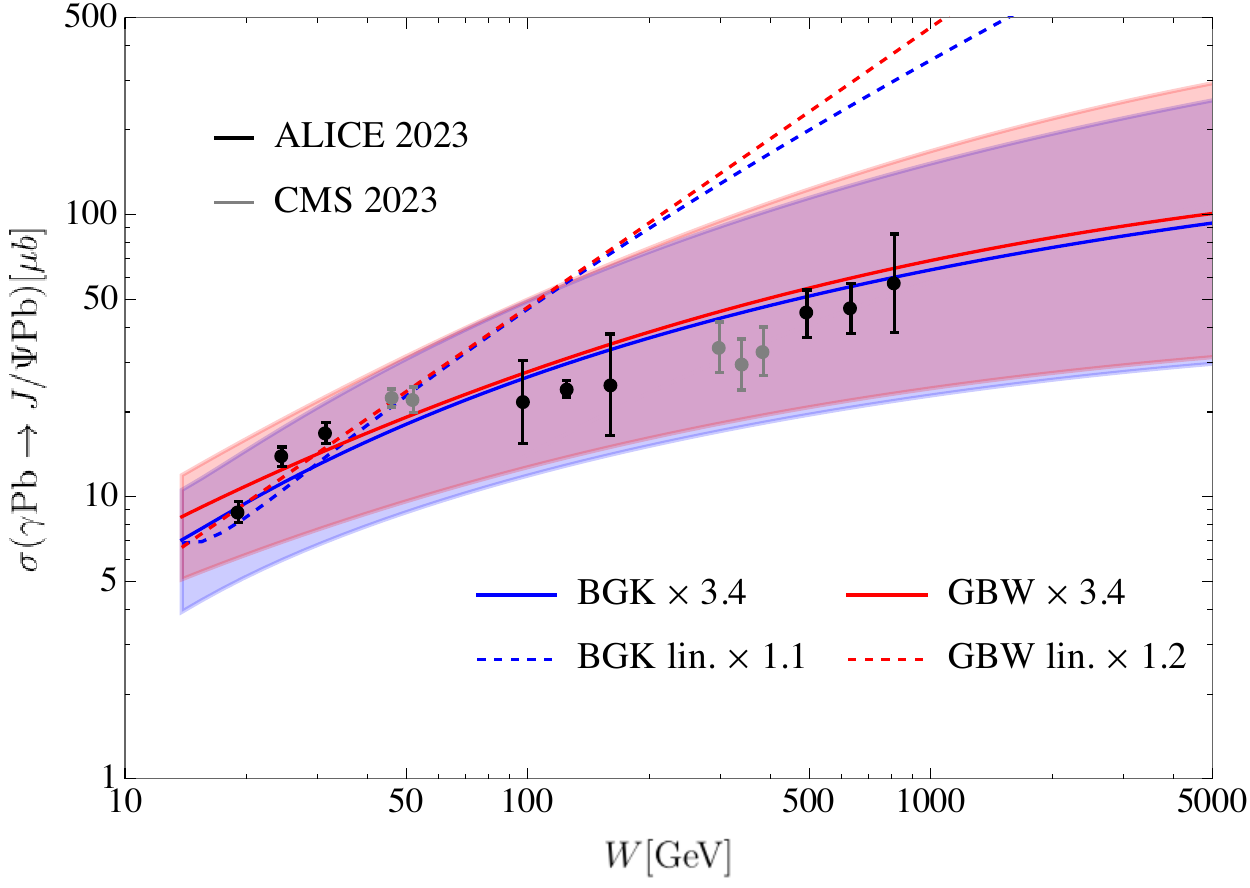}
   \includegraphics[width=.49\textwidth]{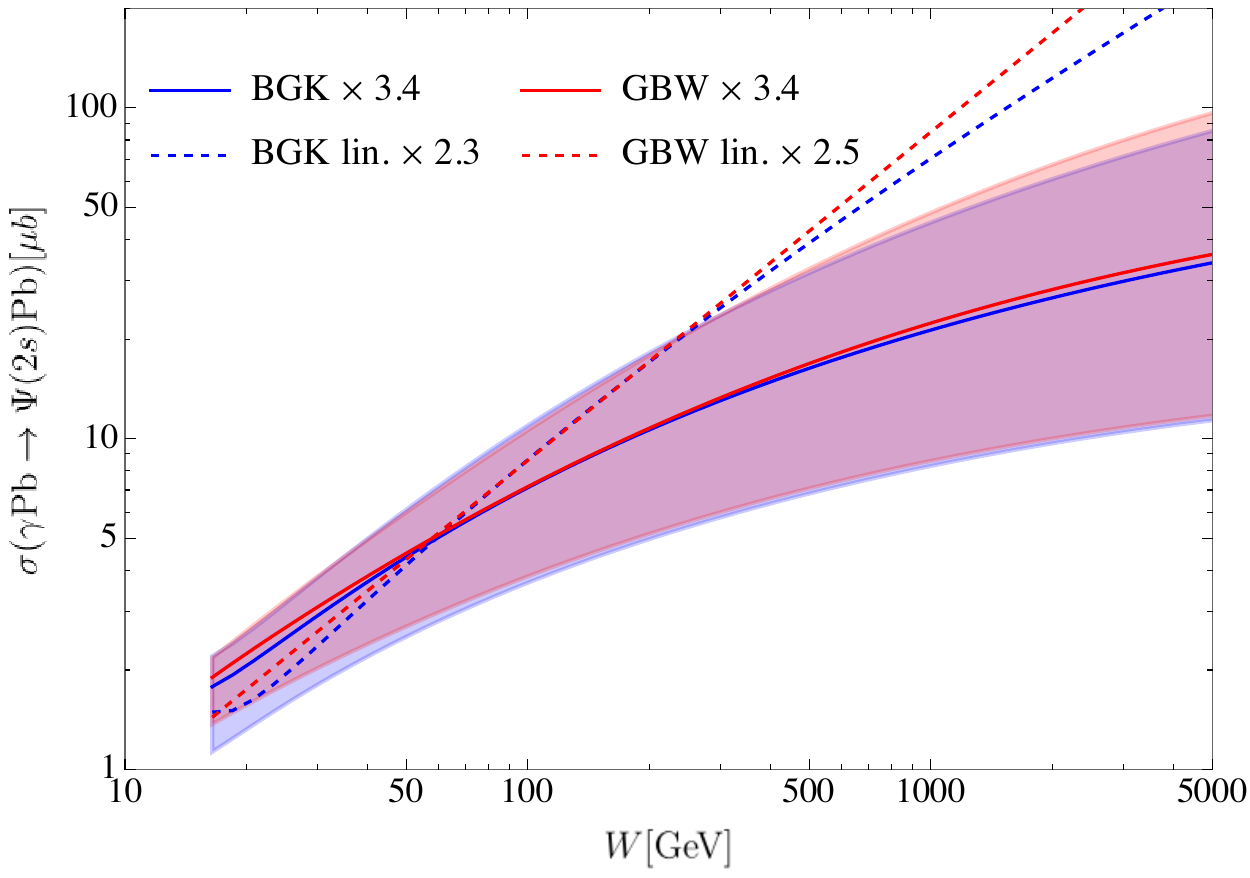}
  \caption{Predictions of the photonuclear production of $J/\Psi$ and $\Psi(2s)$. We further show ALICE \cite{ALICE:2023jgu} and CMS \cite{CMS:2023snh} data. }
  \label{fig:psi2sNUC}
\end{figure}
Results for photoproduction on a lead nucleus are given in
Fig.~\ref{fig:psi2sNUC}.  To provide an estimate of uncertainty
associated with the nuclear scaling of the saturation scale, we vary in Fig.~\ref{fig:psi2sNUC}
the parameter $A^{1/3}$ by a factor of two. We however would like to point out that  this estimate of
uncertainties is not identical to a similar procedure in the case of
the factorization scale since an all-order cross-section would turn
independent of the factorization scale, while the same is not true for
the above variation of the gluon density. It therefore provides an estimate of the uncertainty due to the ad-hoc rescaling Eq.~\eqref{eq:oomph} to arrive at the saturation scale for the dipole nucleus interaction. Note that for the linear result there is no corresponding uncertainty band, since effects related to a variation of the saturation scale are absent in that case. 
 Despite the relatively wide  uncertainty band for the complete model, we find
a clear deviation of data from the linearized description, which
indicates the relevance of non-linear terms. 

\section{Non-linear corrections and the $\Psi(2s)$ over $J/\Psi$ ratio}
\label{sec:observable}

The above results suggest weak non-linear corrections for charmonium production in  photon-proton collisions, while they appear to be sizeable for photoproduction on a large nucleus. In the following, we provide a more systematic exploration of the relevance of these corrections for the description of data and point out how this is  reflected in the ratio of $\Psi(2s)$ and $J/\Psi$ photoproduction cross-sections.

\subsection{Charmonium production and the scaling region}
\label{sec:scaling}

With  
\begin{align}
  \label{eq:QS_GBW}
  Q^{\text{GBW}}_s(x=10^{-6})\simeq 1.58~\text{GeV},
\end{align}
the numerical value of the saturation scale in the proton is close to
the hard scale of the charmonium production cross-section, which is of
the order of the charm mass, $m_c \simeq 1.4$~GeV.  In such a process
one is therefore not sensitive to the saturated region of the dipole
region. Instead one might at most probe the so-called (geometric)
scaling region \cite{Stasto:2000er}, where the dimensionless dipole
cross-section $\sigma_{q\bar{q}}(x, r)/\sigma_0$ is still weak,
$\sigma_{q\bar{q}}(x, r)/\sigma_0 \ll 1$, but already influenced by
the presence of the saturated region
$\sigma_{q\bar{q}}(x, r)/\sigma_0 \sim 1$ \cite{Mueller:2018ned}. This
geometric scaling region can be  defined as the range in dipole
size $r$ in which the dipole cross-section $\sigma_{q\bar{q}}(x, r)$
turns into a function of a single variable,
$\sigma_{q\bar{q}}(x, r) \to \sigma_{q\bar{q}}(r^2 Q_s^2(x))$. Using
properties of solutions to the BK equations, this region can be
estimated due to the following inequality \cite{Mueller:2002zm,
  Munier:2003sj,Munier:2003vc}:
\begin{align}
  \label{eq:estimate}
  1 < \left|\ln \left(r^2 Q_s^2(x) \right) \right| \leq \sqrt{\overline{\alpha}_s \chi_0''(\gamma_0)},
\end{align}
with $\chi_0(\gamma) = 2 \Psi(1) - \Psi(\gamma) - \Psi(1-\gamma)$ the leading order BFKL eigenvalue and $\gamma_0$  given  implicitly  through $\chi_0(\gamma_0)/\gamma_0 = \chi'_0(\gamma_0)$ with $\gamma_0 \simeq 0.627549$.
\begin{figure}[t]
  \centering
  \parbox{.45\textwidth}{
    \includegraphics[width=.45\textwidth]{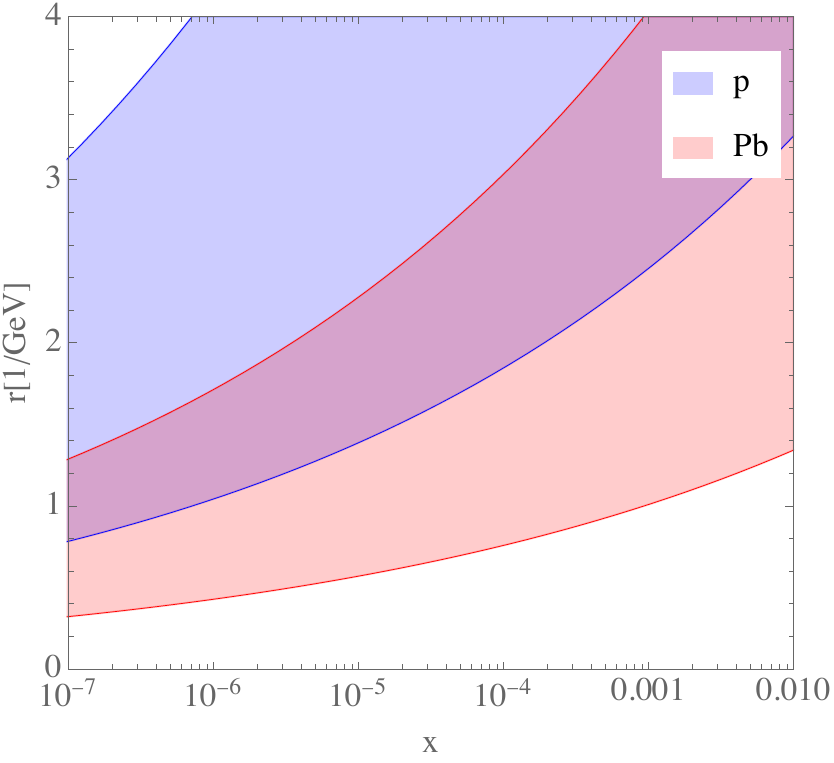}}
\parbox{.45\textwidth}{
   $\qquad$ \includegraphics[width=.45\textwidth]{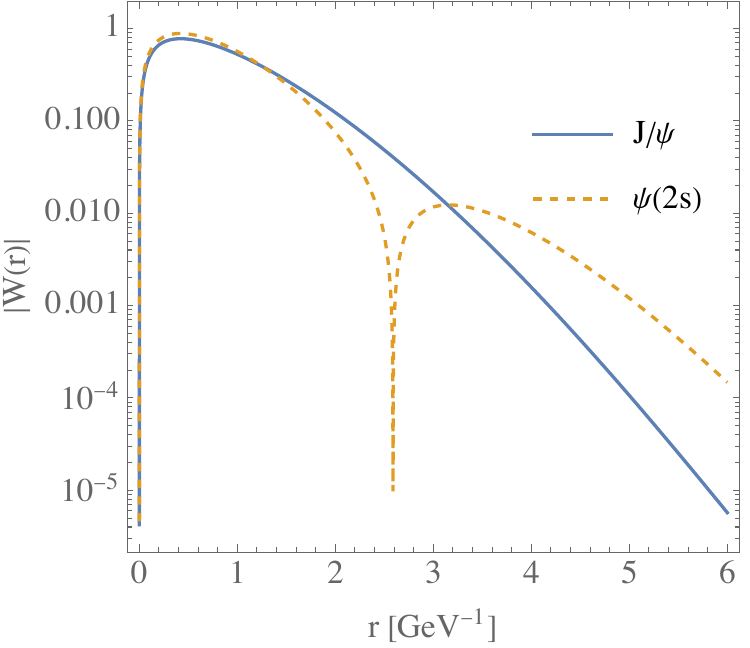}}
   \caption{Left: Estimated geometric scaling region, based on the GBW saturation scale for both proton and lead. Right: Integrated Gaussian wave function overlap for photo-production of vector mesons $J/\Psi$ and $\Psi(2s)$}
  \label{fig:compare}
\end{figure}
The resulting scaling region is illustrated in Fig.~\ref{fig:compare} (left) using the  GBW saturation scale and $\overline{\alpha}_s(m_c) \simeq 0.29$, for both the proton and a lead nucleus.  For the proton, even at the lowest accessible values of $x$, one enters the scaling region only for $r > 1/$GeV, while for a lead nucleus, the scaling regions start already at $r > 0.4/$GeV.   To compare the scaling region to the region in dipole sizes probed for $J/\Psi$ and $\Psi(2s)$ production, we  introduce the normalized wave function overlap
\begin{align}
  \label{eq:Wr}
  W_V(r) & = \frac{r \int_0^1 dz (\Psi^*_V \Psi_T)(r,z)  }{\int dr r \int_0^1 dz (\Psi^*_V \Psi_T)(r,z) },  &
\int_0^\infty dr W(r) & = 1,
\end{align}
see Fig.~\ref{fig:compare}, right; note that $W_{\Psi(2s)}(r) < 0$ for
$r > 2.59/$GeV. To access the relevance of the different regions in
dipole size $r$ for the complete cross-section, we further provide in
Tab.~\ref{tab:WF} the integrated $W_V(r)$ function for different
regions. For the photonuclear production cross-section, the bulk of
the dipole sizes probed in the reaction -- approximately three
quarters -- lies within the geometric scaling region; this explains
the clear imprint of a non-linear energy dependence in the
photoproduction cross-section shown in Fig.~\ref{fig:psi2sNUC}. For
photoproduction on a proton, this contribution is however reduced to
approximately a third, which explains the relatively weak 
non-linear dynamics in Fig.~\ref{fig:full_vs_linear}.
\begin{table}[t]
 \caption{Percentage of $ w_V =\int_{r_{min}}^{r_{max}} dr W_{V}$ for different regions of dipole size $r$. Note that the $W_{\Psi(2s)} < 0$  for  $r >2.59/$GeV.}
  \centering
  \begin{tabular}{r|c|c|c|c}
    & $ 0< r< 0.4/  \text{GeV}$ & $ 0.4< r< 1/  \text{GeV}$ &  $ 1< r< 3/  \text{GeV}$ &$ r> 3/ \text{GeV}$ \\ \hline
$w_{J/\Psi} (\%)$ & 23.6&  40.8  & 34.9  & 0.7 \\
$w_{\Psi(2s)} (\%)$  & 26.7&45.7 & 30.0 & - 1.4
  \end{tabular}
 
  \label{tab:WF}
\end{table}
 We further determine the first moment of the wave function overlap, defined through 
\begin{align}
  \label{eq:moments}
  \langle r \rangle_{V} & =  \int_0^\infty dr\, r \cdot W_V(r), & & \text{with} &  \langle r\rangle_{J/\Psi}&  = 0.89/\text{GeV},  & \langle r\rangle_{\Psi(2s)} & = 0.72/\text{GeV},
\end{align}
 which confirms the observation\cite{Cepila:2019skb, Nemchik:1997xb} that photo-production of a $\Psi(2s)$ is dominated by slightly smaller dipole sizes.  

\subsection{The ratio of the photo-production cross-sections}
\label{sec:start_ratio}

To have access to the geometrical scaling region in the proton, it is therefore necessary to probe observables that are somehow sensitive to dipole sizes  $1< r < 3$/GeV. In this region, both wave function overlaps are still sizeable, while they differ significantly in shape due to the node  of  $W_{\Psi(2s)}$ at $r = 2.59/$GeV. A possibility to gain sensitivity to this region is therefore given by the  ratio of both photo-production cross-sections.   To understand the behavior of this ratio, it is best to study it first for the scenario where non-linear dynamics is absent. For the linearized GBW model,
\begin{align}
  \label{eq:GBWlin}
  \sigma_{q\bar{q}}^{\text{GBW, lin.}}(x, r) & = \sigma_0^{\text{GBW}} r^2 Q_s^2(x)/4, 
\end{align}
 Eq.~\eqref{am-i} turns into
\begin{align}
  \label{eq:6_linearGBW}
  \Im\text{m}\mathcal{A}_T^{\gamma p \to Vp}(x) & = Q_s^2(x) \cdot   \sigma_0^{\text{GBW}}\int d^2 {\bm r} \Sigma(r)  r^2/4, & V = J/\Psi, \Psi(2s).
\end{align}
With the entire $x$ dependence contained in the saturation scale --  which itself is independent of the dipole size --  the  $x$ dependence is independent of the vector meson wave function, and the  GBW saturation scale cancels in the ratio. Leaving aside a possible weak logarithmic dependence on the center of mass energy $W$ due to the slope parameter Eq.~\eqref{eq:18}, one finds a constant dipole cross-section ratio. If, on the other hand, the full non-linear GBW model is inserted,  the  $x$-dependence of the dipole cross-section starts to depend on the size of the dipole and does no longer cancel in the ratio; the impact of these non-linear terms on the ratio will be explored below.

A similar observation holds for the BGK model, where the factorization
scale (and therefore the resulting saturation scale) depends on the
dipole size. Nevertheless in the region of interest corresponding to
dipole sizes $r > 1$/GeV, the factorization scale approaches rapidly
$\mu_0 \simeq 1.32$~GeV and one deals again with a dipole size
independent saturation scale. For $r> 1/$GeV, the $x$-dependence of
the collinear leading order dipole cross-section
Eq.~\eqref{eq:sigcoll} turns therefore $r$-independent and one finds
again a ratio which is approximately $x$ independent, if one sticks
to the linear dipole cross-section. If on the other hand the collinear
dipole cross-section is exponentiated (corresponding to the BGK
model), the ratio of the $\Psi(2s)$ and $J/\Psi$ photo-production
cross-sections turns sensitive to the different shapes of the wave
function overlaps.  To illustrate this point further, we consider the
normalized unintegrated scattering amplitude $a_V(x,r)$,
$V = J/\Psi, \Psi(2s)$, differential in dipole size $r$, which now
includes in addition to Eq.~\eqref{eq:Wr} also effects due to the
dipole cross-section. With
\begin{align}
  \label{eq:dAdr1}
  \Im\text{m} A_V(x) & =  \int_0^\infty dr \, 2 \pi r \Sigma_{\gamma V}(r) \sigma_{q\bar{q}}(x, r),
\end{align}
we define
\begin{align}
  \label{eq:dAdr2}
  a_V(x, r) & = \frac{ 2 \pi r \Sigma_{\gamma V}(r) \sigma_{q\bar{q}}(x, r)}{\Im\text{m} A(x)},
&
\int_0^\infty dr a_V(x, r) & = 1.
\end{align}
\begin{figure}[t]
  \centering
  \includegraphics[width=.3\textwidth]{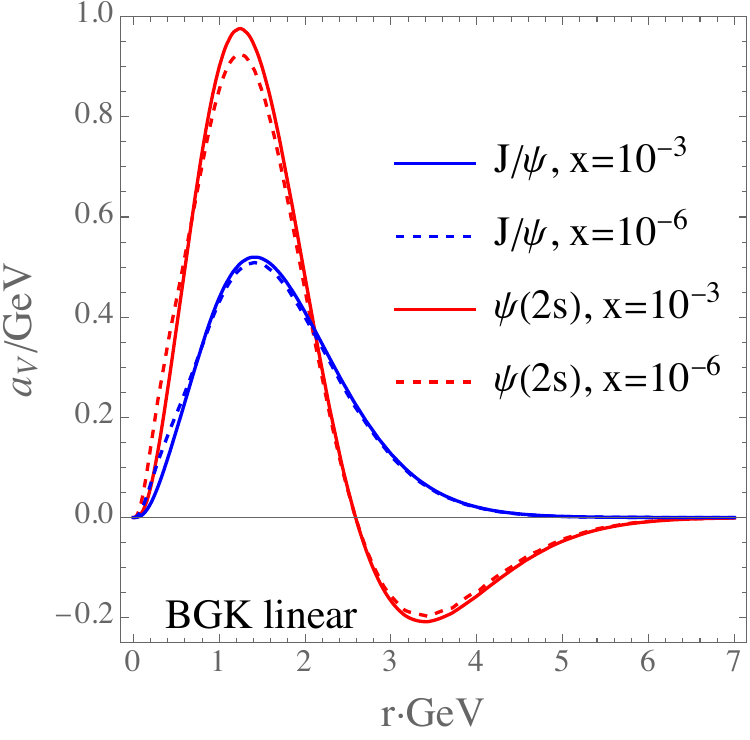}$\quad$ 
  \includegraphics[width=.3\textwidth]{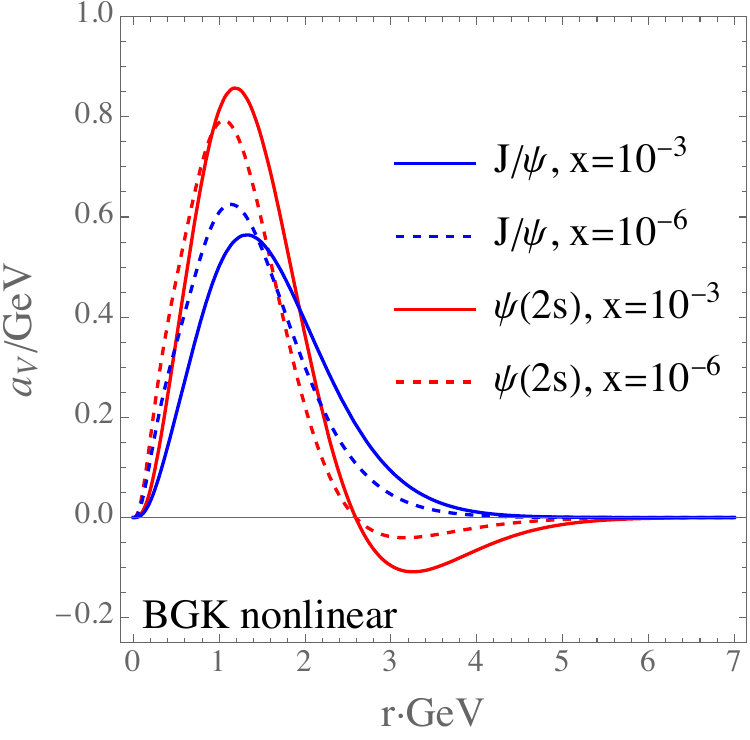} $\quad$  
  \includegraphics[width=.3\textwidth]{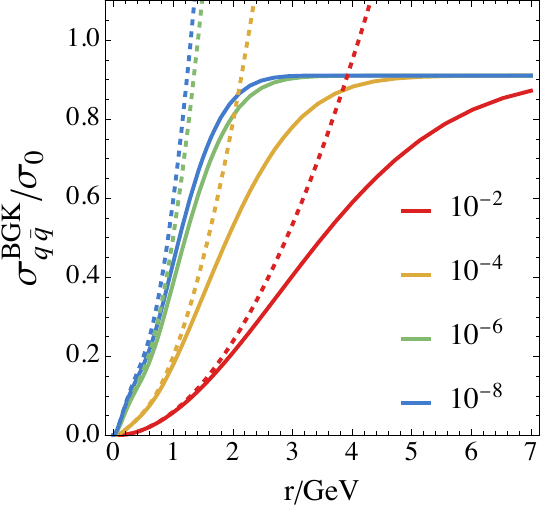}
  \caption{Left: Eq.~\eqref{eq:dAdr2} vs the dipole size $r$ for both the linearized/collinear BGK model Eq.~\eqref{eq:sigBGK}, corresponding to the dipole cross-section to leading order within collinear factorization. Center: the same quantity for the complete BGK model. Both dipole cross-sections are based on the same gluon distribution of \cite{Golec-Biernat:2017lfv}. Right: Evolution of the BGK dipole cross-section towards small $x$, both for the complete (solid) and collinear (dashed).} 
  \label{fig:dadr}
\end{figure}
While the overall value is growing with $x$, the normalized expression Eq.~\eqref{eq:dAdr2} does for the collinear dipole cross-section   not change with $x$, see Fig.~\ref{fig:dadr} (left), leaving aside a small modification in the perturbative region $r< 1/$~GeV which is induced by DGLAP evolution and well under control. For the non-linear case, Fig.~\ref{fig:dadr} (center), we find on the other hand a significant modification both in the region of perturbative dipole sizes $r<1/$GeV as well as in the scaling region. We therefore expect  an increased sensitivity to the geometric scaling region and therefore the presence of high gluon densities. 
\begin{figure}[t]
  \centering
  \includegraphics[height = 4.15cm]{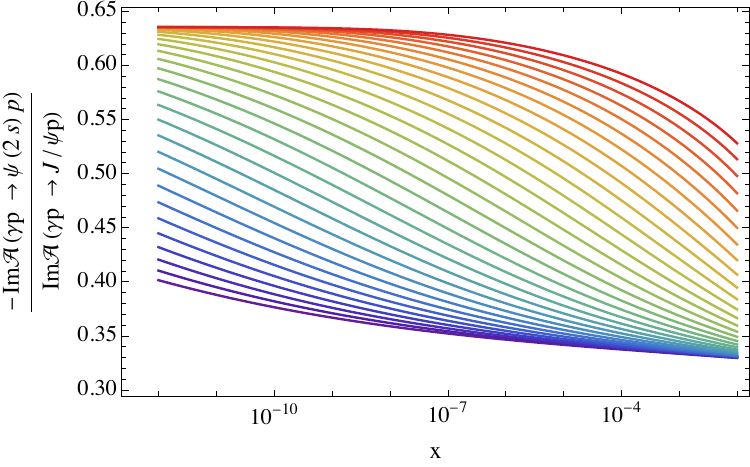} $\quad$
 \includegraphics[height= 4.15cm]{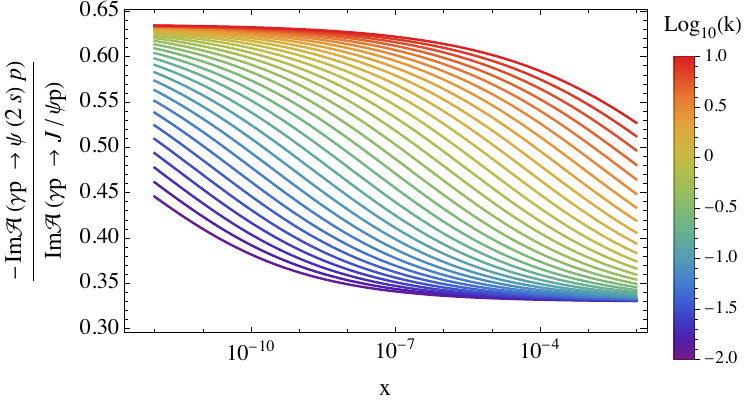}
  \caption{Ratio of the imaginary part of the scattering amplitude for the BGK (left) and GBW (right) model for different values of $k \in [10^{-2}, 10]$}
  \label{fig:ratio_amp}
\end{figure}
In Fig.~\ref{fig:ratio_amp} we explore the effect of different values of $k$ on the behavior of the ratio at amplitude level, down to values of $x=10^{-12}$.  While a linearized dipole cross-section assumes the total absence of non-linear corrections, small values of $k$ probe the scenario where non-linear corrections are in principle present, but very small. In this case,  dipole cross-section unitarizes in the limit $x \to 0$ and $r \to \infty$, while  these corrections are assumed to be numerically very small in the region probed by current experimental facilities. We find that for this scenario, one has indeed an almost constant amplitude ratio, as seen for the linearized case. While the GBW model yields an essentially constant ratio for small $k$ and/or intermediate $x$, the BGK model predicts even for that region a slight growth, which is however substantially weaker than the growth predicted in the presence of sizeable non-linear corrections.  Independent of the strength of non-linear corrections, i.e., independent of the value of $k$,  we find that there is always a region in $x$, where the ratio starts to grow. The onset of this growth is strongly correlated with the value of the parameter $k$, i.e., for large values of $k$ -- corresponding to higher gluon densities -- the growth starts for larger values of $x$ than for smaller values of $x$. 

 We further observe  that the growth of the ratio eventually saturates once sufficiently high densities are reached -- either through a sufficiently large $k$ parameter or through continuation to ultra-small $x$ values. 
This behavior can be understood from both the behavior of the dipole cross-section for ultra-small $x$, see Fig.~\ref{fig:dadr}(right) and the normalized wave function overlap Eq.~\eqref{eq:Wr},  Fig.~\ref{fig:compare}(right): Turning to lower and lower values $x$, the active region of the dipole,  i.e. the region changing  shape with decreasing $x$, is moving towards smaller and smaller dipole sizes. Eventually one reaches values of the dipole size $r$, where $W_V(r)$ coincides for both vector mesons and the ratio approaches again a constant value, which is related to the difference in the  overall normalization. 

 While values $x < 10^{-7}$ are essentially impossible to observe at the LHC, such a behavior would be already visible at LHC, if the saturation scale would be twice as large  ($k=4$) as extracted from HERA data, which is approximately realized for photonuclear production on Pb, since $A^{1/3}_{\text{Pb}} \simeq 5.92$.

\subsection{Comparison to collinear factorization}
\label{sec:coll_fac}

The above result suggests -- in line with the observation made in \cite{Hentschinski:2020yfm} -- that a rising cross-section ratio yields a clear indicator for the presence of non-linear effects. It is therefore natural to investigate the ratio also from the point of view of collinear factorization.  As far as the determination of the  dipole cross-section within collinear factorization is concerned, we already addressed this question when considering the linearized BGK model:  the scattering amplitude is approximately constant with decreasing values of $x$. There are however different scenarios possible, if  one  determines the entire cross-section within collinear factorization, making use of non-relativistic QCD to describe the photon to vector meson transition. Within such a framework, the leading order QCD cross-section is given by  \cite{Ryskin:1992ui}, see also \cite{Flett:2020duk,Jones:2013eda,Jones:2013pga},
\begin{align}
  \label{eq:collXse}
  \frac{d\sigma}{dt}(\gamma p \to Vp)\bigg|_{t=0} & = \frac{\Gamma^{V}_{ee} M_V^3 \pi^3}{48 \alpha_{e.m.}} \left[\frac{\alpha_s(\mu^2)}{\bar{Q}^4}  xg\left(x,\mu^2\right)\right]^2;
\end{align}
for the next-to-leading order result see  \cite{Flett:2021ghh, Eskola:2022vpi, Eskola:2022vaf}. Here
$\mu$ denotes the renormalization and factorization scale, while
$\bar{Q}^2 = m_c^2$ for the case of a real photon with
$\Gamma_{ee}^{J/\Psi} = 5.55\cdot 10^{-6}$~GeV and
$\Gamma_{ee}^{\Psi(2s)} = 2.33\cdot 10^{-6}$~GeV the electric width of
the vector meson. Such a description is based on the assumption that
the wave function overlap of both vector mesons is dominated by small
dipole sizes, which is in accordance with the observation made in
Tab.~\ref{tab:WF}, at least as far as photoproduction on a proton is concerned. Since the hard scale  is of the order of the charm mass, it is natural to associate the latter with the
factorization and renormalization scale of the above cross-section. In
such a case, the ratio of $\Psi(2s)$ and $J/\Psi$ photo-production cross-sections is trivially constant with decreasing $x$, since the
gluon distribution in  Eq.~\eqref{eq:collXse}  cancels in the ratio. 

 As an alternative to this
choice one can also argue along the  observation made at the end of
Sec.~\ref{sec:scaling}: the first moment of the wave function overlap
is smaller for  $\Psi(2s)$ than for $J/\Psi$. This then  suggests  that the factorization scale of the $\Psi(2s)$ might be  slightly larger than the one of the  $J/\Psi$. Indeed, this is a common choice in the
literature, see  for instance \cite{Jones:2013eda, Jones:2013pga}, where one uses  $\mu =   M_V/2$.  Since the growth of the gluon distribution with decreasing $x$ is stronger for larger factorization scales, one therefore finds also a  rising cross-section ratio within collinear factorization. Collinear factorization is therefore  able to accommodate  a rise through a corresponding {\it choice} of the factorization scale. The main difference is that collinear factorization is able to accommodate a rise of the cross-section, while a non-linear dipole cross-section predicts such a rise. Within collinear factorization,  a constant or -- if desired -- even  falling ratio with decreasing $x$ is equally possible. The conventional procedure to reduce such a scale ambiguity requires the determination of higher order perturbative corrections. While NLO corrections are known, see e.g.  \cite{Flett:2021ghh, Eskola:2022vpi, Eskola:2022vaf}, the perturbative corrections which control logarithms in the factorization scale are naturally proportional to the leading order DGLAP gluon-to-gluon and quark-to-gluon splitting function and therefore very large in the low $x$ region; the scale uncertainty is therefore not considerably reduced. It is therefore probably necessary to  either determine corrections beyond NLO  or some to implement a low $x$ resummation within the collinear calculation. We believe it would be interesting to investigate this problem further along the lines of \cite{Lansberg:2021vie, Lansberg:2023kzf}, which addresses a related problem in the context of inclusive charmonium production.

Within the complete dipole models, the rise of the ratio is, on the other hand, a direct consequence of the above discussed low $x$ dynamics, {\it i.e.}, the rise can be understood as a prediction of low $x$ dynamics, which is directly related to non-linear low $x$ dynamics in the dipole cross-sections. Observation of a constant or very slowly growing ratio would be a strong hint towards a significant delay of the onset of non-linear QCD dynamics in the currently available data set. If one would extend, on the other hand, the data set beyond $x=10^{-6}$, the growth of the ratio would start to slow down if caused by non-linear dynamics. In contrast, collinear factorization would not be able to accommodate such a feature. Indeed, the onset of such a slowdown might be already observable for photonuclear reactions, where gluon densities are enhanced by $A^{1/3}$.

\subsection{Results at cross-section level }
\label{sec:num}

\begin{figure}[t]
  \centering
   \includegraphics[width=.45\textwidth]{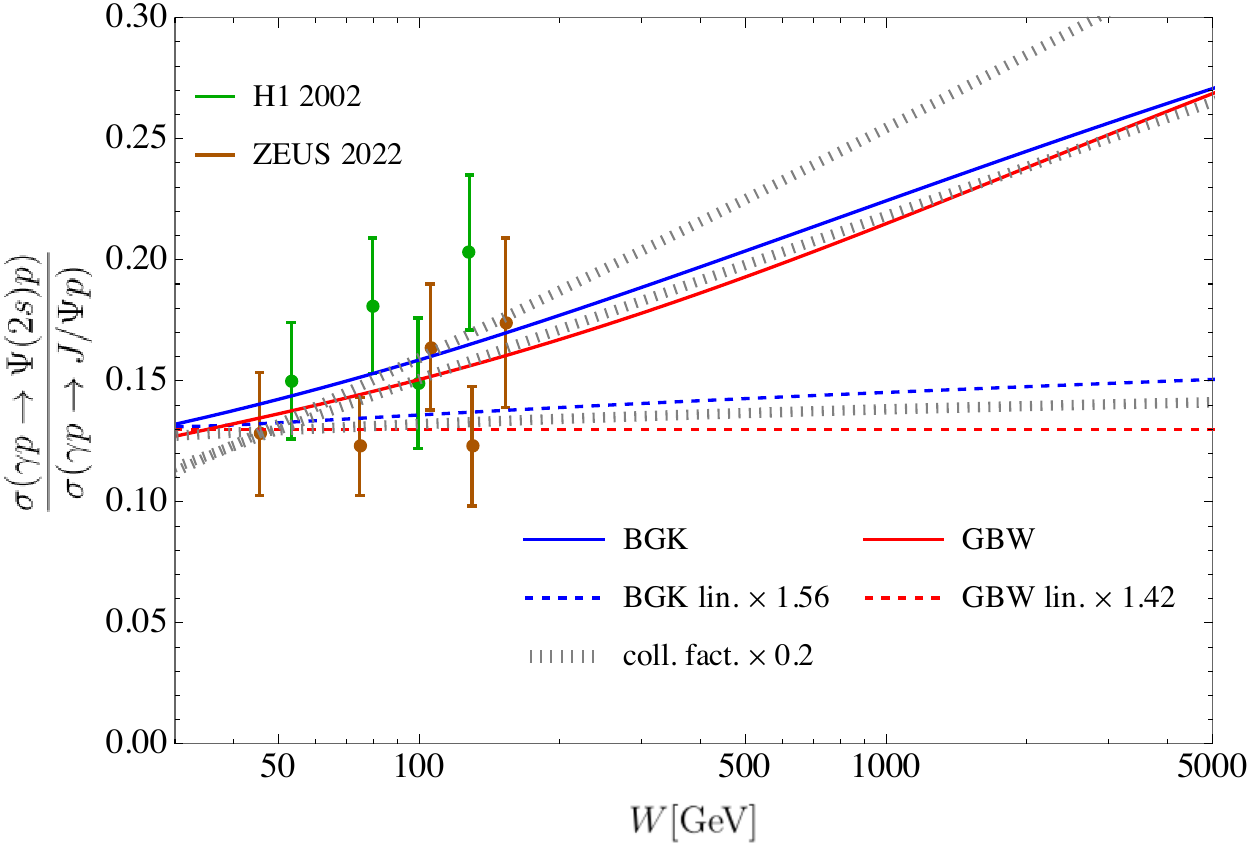}
 \includegraphics[width=.45\textwidth]{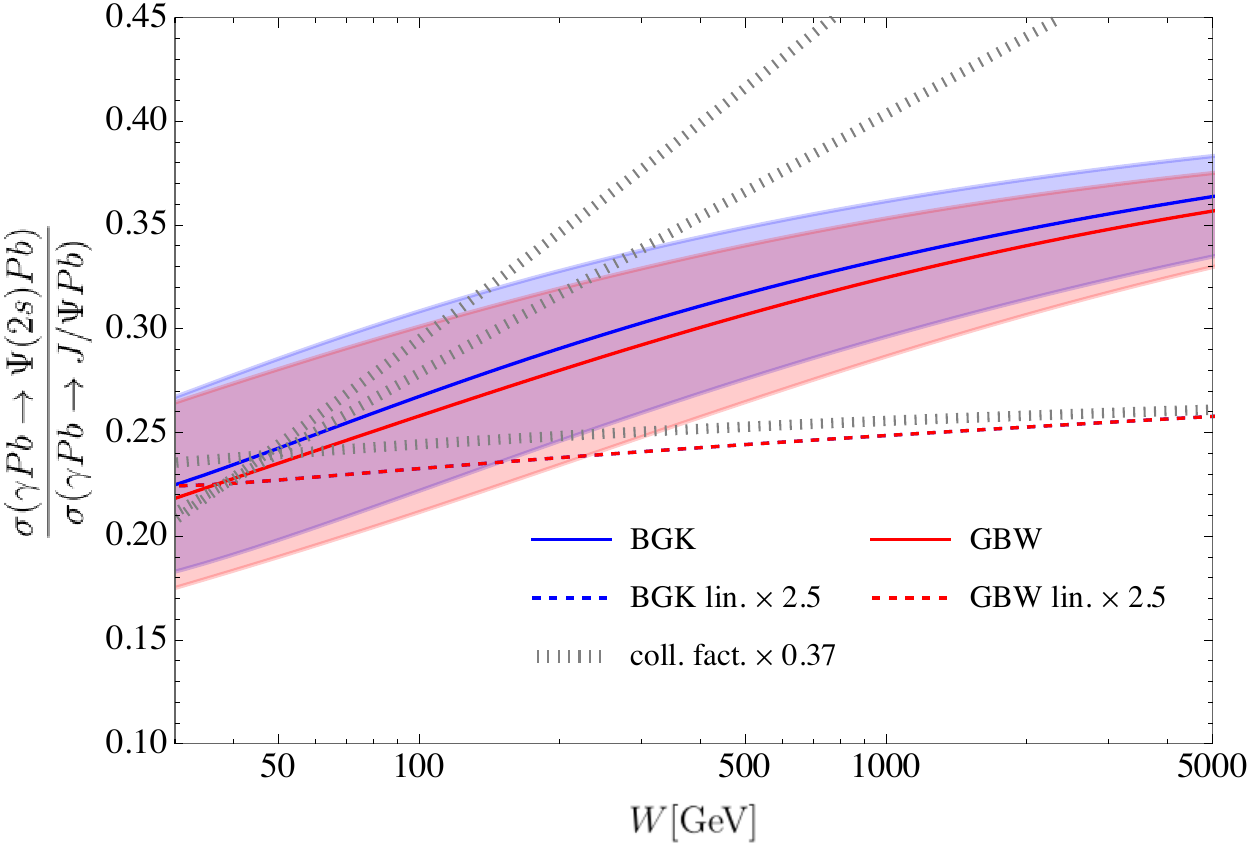}
   \caption{Ratio of photoproduction cross-sections both for the proton (left) and lead (right). For the proton we further depict ZEUS \cite{ZEUS:2022sxn} and H1 \cite{H1:2002yab} data }
  \label{fig:ratio_pPb}
\end{figure}

In Fig.~\ref{fig:ratio_pPb}, we finally provide our predictions for the ratio at the cross-section level. We present both results for exclusive photoproduction on a proton, for which we compare to recent ZEUS as well as H1 data and predictions for exclusive photonuclear production on a lead nucleus. As expected from the above study on the amplitude level,  a linearized dipole cross-section provides, in both cases, an almost constant cross-section ratio,  both for the photoproduction on a proton and on lead. For the complete dipole models Eq.~\eqref{eq:sigBGK} and Eq.~\eqref{eq:sigGBW}, we find a ratio with an approximately linear growth with $\ln W/$GeV.   We also provide predictions based on collinear factorization, where we use the inclusive gluon PDF of the BGK fit as the underlying parton distribution function. As expected, we find a constant ratio if the factorization scale is chosen as the charm mass $m_c = 1.4$~GeV for both production cross-sections.
In contrast, we find rising ratios if we introduce a hierarchy between the $\Psi(2s)$ factorization scales and the $J/\Psi$ production cross-sections. In particular, the choice $\mu = M_V/2$, $V= J/\Psi, \Psi(2s)$ yields a cross-section ratio which produces a rise similar to the complete saturation models. While the $K$-factor appears to be large in the case of collinear factorization, we note that they approach unity if one chooses different charm masses for the  $\Psi(2s)$ and the $J/\Psi$, i.e., through setting $M_c = M_V/2$. The depicted results use, on the other hand, a universal value of $m_c = 1.4$~GeV, which cancels for the ratio. We further find that the ratio aligns for the complete dipole models well at low center of mass energies with existing HERA data without the need to introduce additional $ K$ factors. This points towards another advantage of the cross-section ratios since they benefit from the cancellation of various sources of uncertainties for the theoretical predictions, while a similar reduction in uncertainties can be expected for experimental results. 

For photonuclear production, we first note that within our current treatment, all linear descriptions, i.e., linearized dipole models and collinear factorization, coincide with the result presented for the proton case since within or treatment, only the overall normalization is changed for the nuclear case.
 For the completely unitarized dipole models  Eq.~\eqref{eq:sigBGK} and Eq.~\eqref{eq:sigGBW}, with nuclear effects implemented according to  Eq.~\eqref{eq:qqb_A},  we find that the cross-section ratio rises approximately linearly with  $\ln W/$GeV, while this growth slows down with increasing values of $W$, indicating that we are entering the region, where the growth of the ratio starts to saturate. Albeit such a behavior appears to be difficult to identify in experimental data, this would be certainly a feature which collinear factorization could not reproduce, at least within the setup chosen in this paper.

\section{Conclusions}
\label{sec:concl}
In this paper we provided a detailed study of the proposal made in \cite{Hentschinski:2020yfm}, i.e., the claim that a rising ratio of photoproduction cross-sections for $\Psi(2s)$ and $J/\Psi$ can serve as an indicator for the presence of non-linear low $x$ dynamics in the proton and/or a large nucleus. Our study was based on dipole cross-sections provided by the  GBW and BGK dipole models, which essentially exponentiate linear cross-sections to explore the effects of non-linear low $x$ dynamics. As a first result, we found that the observation made in   \cite{Hentschinski:2020yfm} holds if gluon distributions subject to linear NLO BFKL evolution and non-linear BK evolution are replaced with linearized and unitarized dipole models: in the absence of non-linear low $x$ dynamics, the ratio of both cross-sections is approximately constant, whereas the ratio rises for the fully exponentiated dipole model. This observation has a simple explanation in terms of the convolution of dipole cross-section and wave function overlap: If the shape of the dipole cross-section changes with $x$ in a region where wave function overlaps for the transition photon to $\Psi(2s)$ and $J/\Psi$ differ, one obtains a growing ratio. For the linearized case, this phenomenon is absent since, in that case, only the normalization changes with $x$. At the same time, the shape in dipole size is approximately $x$ independent, leaving aside small corrections induced by DGLAP evolution. 

The behavior of the ratio is therefore characteristically different from the behavior of the photo-production cross-section themselves, at least as far as photoproduction on a proton is concerned: While for the photo-production cross-section the dominant part of the wave function overlap lies outside of the so-called geometric scaling region of the dipole cross-section, the region where both wave function overlaps differ is  placed at dipole sizes inside the geometric scaling region. For photoproduction of $\Psi(2s)$ and $J/\Psi$ on a lead nucleus,  the gluon density is  enhanced by a factor of order $A_{\text{Pb}}^{1/3}\simeq 5.92$, which extends the geometric scaling region  deep into the region of dipole separations, where the  $\Psi(2s)$ and $J/\Psi$ wave function overlap  receive their dominant contribution. As a consequence one finds a clear difference between photo-production cross-sections based on linearized and complete dipole cross-sections, where the latter are clearly preferred by data.

 This observation is in our opinion a clear validation of the underlying concepts of gluon saturation: For  $J/\Psi$ production on a proton, the difference between descriptions based on unitarized and linear dipole cross-section is small  and most likely lies within the uncertainties of light-front wave functions and employed saturation models. For   $J/\Psi$ production on a lead nucleus, the differences are on the other hand substantial and the non-linear description is clearly preferred by data. We believe that it would be highly beneficial to complement this picture through a measurement of the cross-section ratio for both photoproduction on a proton and on a nucleus.

Another direction of research which we did not explore in this paper, is the description of the cross-section ratio in Deep Inelastic Scattering, which opens the possibility to study in addition the dependence on photon virtuality. We plan to explore this possibility in the future. While there exist already HERA data on the the cross-section ratio for photoproduction on a proton \cite{ZEUS:2022sxn} which allows to study the dependence on photon virtuality, the future Electron Ion Collider  \cite{Accardi:2012qut, AbdulKhalek:2021gbh,AbdulKhalek:2022hcn} will allow enable to combine this with a  study of nuclear effects.

\section*{Acknowledgments}

We acknowledge support by Consejo Nacional de Ciencia y Tecnolog{\'\i}a grant number A1 S-43940 (CONACYT-SEP Ciencias B{\'a}sicas). We would like to thank Heikki M\"antysaari for pointing out to us the incorrect use of incoherent instead of coherent photonuclear  ALICE data for the determination of the nuclear diffractive slope in our initial manuscript. 

\bibliographystyle{hunsrt} 
\bibliography{paper}

\begin{thebibliography}{10}

\bibitem{Kuraev:1977fs}
E.A. Kuraev, L.N. Lipatov, and Victor~S. Fadin.
\newblock {The Pomeranchuk Singularity in Nonabelian Gauge Theories}.
\newblock {\em Sov. Phys. JETP}, 45:199--204, 1977.

\bibitem{Kuraev:1976ge}
E.~A. Kuraev, L.~N. Lipatov, and Victor~S. Fadin.
\newblock {Multi - Reggeon Processes in the Yang-Mills Theory}.
\newblock {\em Sov. Phys. JETP}, 44:443--450, 1976.

\bibitem{Balitsky:1978ic}
I.I. Balitsky and L.N. Lipatov.
\newblock {The Pomeranchuk Singularity in Quantum Chromodynamics}.
\newblock {\em Sov. J. Nucl. Phys.}, 28:822--829, 1978.

\bibitem{Gribov:1984tu}
L.V. Gribov, E.M. Levin, and M.G. Ryskin.
\newblock {Semihard Processes in QCD}.
\newblock {\em Phys. Rept.}, 100:1--150, 1983.

\bibitem{Gelis:2010nm}
Francois Gelis, Edmond Iancu, Jamal Jalilian-Marian, and Raju Venugopalan.
\newblock {The Color Glass Condensate}.
\newblock {\em Ann. Rev. Nucl. Part. Sci.}, 60:463--489, 2010, 1002.0333.

\bibitem{Morreale:2021pnn}
Astrid Morreale and Farid Salazar.
\newblock {Mining for Gluon Saturation at Colliders}.
\newblock {\em Universe}, 7(8):312, 2021, 2108.08254.

\bibitem{Balitsky:1995ub}
I.~Balitsky.
\newblock {Operator expansion for high-energy scattering}.
\newblock {\em Nucl. Phys. B}, 463:99--160, 1996, hep-ph/9509348.

\bibitem{Jalilian-Marian:1997ubg}
Jamal Jalilian-Marian, Alex Kovner, and Heribert Weigert.
\newblock {The Wilson renormalization group for low x physics: Gluon evolution
  at finite parton density}.
\newblock {\em Phys. Rev. D}, 59:014015, 1998, hep-ph/9709432.

\bibitem{Kovchegov:1999yj}
Yuri~V. Kovchegov.
\newblock {Small x F(2) structure function of a nucleus including multiple
  pomeron exchanges}.
\newblock {\em Phys. Rev. D}, 60:034008, 1999, hep-ph/9901281.

\bibitem{Iancu:2000hn}
Edmond Iancu, Andrei Leonidov, and Larry~D. McLerran.
\newblock {Nonlinear gluon evolution in the color glass condensate. 1.}
\newblock {\em Nucl. Phys. A}, 692:583--645, 2001, hep-ph/0011241.

\bibitem{Weigert:2000gi}
Heribert Weigert.
\newblock {Unitarity at small Bjorken x}.
\newblock {\em Nucl. Phys. A}, 703:823--860, 2002, hep-ph/0004044.

\bibitem{Iancu:2001ad}
Edmond Iancu, Andrei Leonidov, and Larry~D. McLerran.
\newblock {The Renormalization group equation for the color glass condensate}.
\newblock {\em Phys. Lett. B}, 510:133--144, 2001, hep-ph/0102009.

\bibitem{Ferreiro:2001qy}
Elena Ferreiro, Edmond Iancu, Andrei Leonidov, and Larry McLerran.
\newblock {Nonlinear gluon evolution in the color glass condensate. 2.}
\newblock {\em Nucl. Phys. A}, 703:489--538, 2002, hep-ph/0109115.

\bibitem{Celiberto:2023fzz}
Francesco~Giovanni Celiberto.
\newblock {Vector Quarkonia at the LHC with Jethad: A High-Energy Viewpoint}.
\newblock {\em Universe}, 9(7):324, 2023, 2305.14295.

\bibitem{Chapon:2020heu}
Emilien Chapon et~al.
\newblock {Prospects for quarkonium studies at the high-luminosity LHC}.
\newblock {\em Prog. Part. Nucl. Phys.}, 122:103906, 2022, 2012.14161.

\bibitem{Cepila:2018faq}
J.~Cepila, J.G. Contreras, and M.~Matas.
\newblock {Collinearly improved kernel suppresses Coulomb tails in the
  impact-parameter dependent Balitsky-Kovchegov evolution}.
\newblock {\em Phys. Rev. D}, 99(5):051502, 2019, 1812.02548.

\bibitem{Krelina:2019gee}
M.~Krelina, V.~P. Goncalves, and J.~Cepila.
\newblock {Coherent and incoherent vector meson electroproduction in the future
  electron-ion colliders: the hot-spot predictions}.
\newblock {\em Nucl. Phys. A}, 989:187--200, 2019, 1905.06759.

\bibitem{Klein:2019qfb}
Spencer~R. Klein and Heikki M\"antysaari.
\newblock {Imaging the nucleus with high-energy photons}.
\newblock {\em Nature Rev. Phys.}, 1(11):662--674, 2019, 1910.10858.

\bibitem{Kopeliovich:2020has}
B.~Z. Kopeliovich, M.~Krelina, J.~Nemchik, and I.~K. Potashnikova.
\newblock {Ultraperipheral nuclear collisions as a source of heavy quarkonia}.
\newblock {\em Phys. Rev. D}, 107(5):054005, 2023, 2008.05116.

\bibitem{Bendova:2020hbb}
D.~Bendova, J.~Cepila, J.~G. Contreras, and M.~Matas.
\newblock {Photonuclear $J/\psi$ production at the LHC: Proton-based versus
  nuclear dipole scattering amplitudes}.
\newblock {\em Phys. Lett. B}, 817:136306, 2021, 2006.12980.

\bibitem{Jenkovszky:2021sis}
Laszlo Jenkovszky, Vladyslav Libov, and Magno V.~T. Machado.
\newblock {The reggeometric pomeron and exclusive production of
  J/\ensuremath{\psi} and \ensuremath{\psi}(2S) in ultraperipheral collisions
  at the LHC}.
\newblock {\em Phys. Lett. B}, 824:136836, 2022, 2111.13389.

\bibitem{Flett:2021xsl}
Christopher~Alexander Flett.
\newblock {\em {Exclusive Observables to NLO and Low x PDF Phenomenology at the
  LHC}}.
\newblock PhD thesis, U. Liverpool (main), U. Liverpool (main), 2021.

\bibitem{Mantysaari:2021ryb}
Heikki M\"antysaari and Jani Penttala.
\newblock {Exclusive heavy vector meson production at next-to-leading order in
  the dipole picture}.
\newblock {\em Phys. Lett. B}, 823:136723, 2021, 2104.02349.

\bibitem{Mantysaari:2022kdm}
Heikki M\"antysaari and Jani Penttala.
\newblock {Complete calculation of exclusive heavy vector meson production at
  next-to-leading order in the dipole picture}.
\newblock {\em JHEP}, 08:247, 2022, 2204.14031.

\bibitem{Goncalves:2022ret}
Victor~P. Goncalves, Bruno~D. Moreira, and Luana Santana.
\newblock {Exclusive \ensuremath{\rho} and J/\ensuremath{\Psi} photoproduction
  in ultraperipheral pO and OO collisions at energies available at the CERN
  Large Hadron Collider}.
\newblock {\em Phys. Rev. C}, 107(5):055205, 2023, 2210.11911.

\bibitem{Wang:2022vhr}
Xiao-Yun Wang, Fancong Zeng, and Quanjin Wang.
\newblock {Systematic analysis of the proton mass radius based on
  photoproduction of vector charmoniums}.
\newblock {\em Phys. Rev. D}, 105(9):096033, 2022, 2204.07294.

\bibitem{Klein:2020nvu}
Spencer Klein et~al.
\newblock {New opportunities at the photon energy frontier}.
\newblock 9 2020, 2009.03838.

\bibitem{Bylinkin:2022wkm}
Alexander Bylinkin, Joakim Nystrand, and Daniel Tapia~Takaki.
\newblock {Vector meson photoproduction in UPCs with FoCal}.
\newblock 11 2022, 2211.16107.

\bibitem{ALICE:2023fov}
{Physics of the ALICE Forward Calorimeter upgrade}.
\newblock 2023.

\bibitem{Amoroso:2022eow}
S.~Amoroso et~al.
\newblock {Snowmass 2021 Whitepaper: Proton Structure at the Precision
  Frontier}.
\newblock {\em Acta Phys. Polon. B}, 53(12):12--A1, 2022, 2203.13923.

\bibitem{Hentschinski:2022xnd}
Martin Hentschinski et~al.
\newblock {White Paper on Forward Physics, BFKL, Saturation Physics and
  Diffraction}.
\newblock {\em Acta Phys. Polon. B}, 54(3):3--A2, 2023, 2203.08129.

\bibitem{Frankfurt:2022jns}
L.~Frankfurt, V.~Guzey, A.~Stasto, and M.~Strikman.
\newblock {Selected topics in diffraction with protons and nuclei: past,
  present, and future}.
\newblock {\em Rept. Prog. Phys.}, 85(12):126301, 2022, 2203.12289.

\bibitem{Hentschinski:2020yfm}
Martin Hentschinski and Emilio Padr\'on~Molina.
\newblock {Exclusive $J/\Psi$ and $\Psi(2s)$ photo-production as a probe of QCD
  low $x$ evolution equations}.
\newblock {\em Phys. Rev. D}, 103(7):074008, 2021, 2011.02640.

\bibitem{Garcia:2019tne}
A.~Arroyo~Garcia, M.~Hentschinski, and K.~Kutak.
\newblock {QCD evolution based evidence for the onset of gluon saturation in
  exclusive photo-production of vector mesons}.
\newblock {\em Phys. Lett. B}, 795:569--575, 2019, 1904.04394.

\bibitem{Bautista:2016xnp}
I.~Bautista, A.~Fernandez~Tellez, and Martin Hentschinski.
\newblock {BFKL evolution and the growth with energy of exclusive $J/\psi$ and
  $\Upsilon$ photoproduction cross sections}.
\newblock {\em Phys. Rev. D}, 94(5):054002, 2016, 1607.05203.

\bibitem{Hentschinski:2012kr}
Martin Hentschinski, Agust\'\i{}n Sabio~Vera, and Clara Salas.
\newblock {Hard to Soft Pomeron Transition in Small-x Deep Inelastic Scattering
  Data Using Optimal Renormalization}.
\newblock {\em Phys. Rev. Lett.}, 110(4):041601, 2013, 1209.1353.

\bibitem{Hentschinski:2013id}
Martin Hentschinski, Agustin Sabio~Vera, and Clara Salas.
\newblock {$F_2$ and $F_L$ at small $x$ using a collinearly improved BFKL
  resummation}.
\newblock {\em Phys. Rev. D}, 87(7):076005, 2013, 1301.5283.

\bibitem{Chachamis:2015ona}
Grigorios Chachamis, Michal De\'ak, Martin Hentschinski, Germ\'an Rodrigo, and
  Agust\'\i{}n Sabio~Vera.
\newblock {Single bottom quark production in k$_{\perp}$-factorisation}.
\newblock {\em JHEP}, 09:123, 2015, 1507.05778.

\bibitem{Kutak:2012rf}
Krzysztof Kutak and Sebastian Sapeta.
\newblock {Gluon saturation in dijet production in p-Pb collisions at Large
  Hadron Collider}.
\newblock {\em Phys. Rev. D}, 86:094043, 2012, 1205.5035.

\bibitem{Cepila:2020uxc}
Jan Cepila and Marek Matas.
\newblock {Contribution of the non-linear term in the
  Balitsky\textendash{}Kovchegov equation to the nuclear structure functions}.
\newblock {\em Eur. Phys. J. A}, 56(9):232, 2020, 2006.16136.

\bibitem{Nemchik:1997xb}
J.~Nemchik, Nikolai~N. Nikolaev, E.~Predazzi, B.~G. Zakharov, and V.~R. Zoller.
\newblock {The Diffraction cone for exclusive vector meson production in deep
  inelastic scattering}.
\newblock {\em J. Exp. Theor. Phys.}, 86:1054--1073, 1998, hep-ph/9712469.

\bibitem{Cepila:2019skb}
Jan Cepila, Jan Nemchik, Michal Krelina, and Roman Pasechnik.
\newblock {Theoretical uncertainties in exclusive electroproduction of S-wave
  heavy quarkonia}.
\newblock {\em Eur. Phys. J. C}, 79(6):495, 2019, 1901.02664.

\bibitem{STAR:2006dgg}
J.~Adams et~al.
\newblock {Forward neutral pion production in p+p and d+Au collisions at
  s(NN)**(1/2) = 200-GeV}.
\newblock {\em Phys. Rev. Lett.}, 97:152302, 2006, nucl-ex/0602011.

\bibitem{Dominguez:2011cy}
F.~Dominguez, D.~E. Kharzeev, E.~M. Levin, A.~H. Mueller, and K.~Tuchin.
\newblock {Gluon saturation effects on the color singlet J/$\psi$ production in
  high energy dA and AA collisions}.
\newblock {\em Phys. Lett. B}, 710:182--187, 2012, 1109.1250.

\bibitem{Golec-Biernat:1998zce}
Krzysztof~J. Golec-Biernat and M.~Wusthoff.
\newblock {Saturation effects in deep inelastic scattering at low Q**2 and its
  implications on diffraction}.
\newblock {\em Phys. Rev. D}, 59:014017, 1998, hep-ph/9807513.

\bibitem{Bartels:2002cj}
J.~Bartels, Krzysztof~J. Golec-Biernat, and H.~Kowalski.
\newblock {A modification of the saturation model: DGLAP evolution}.
\newblock {\em Phys. Rev. D}, 66:014001, 2002, hep-ph/0203258.

\bibitem{ALICE:2023jgu}
Shreyasi Acharya et~al.
\newblock {Energy dependence of coherent photonuclear production of J/$\psi$
  mesons in ultra-peripheral Pb-Pb collisions at $\sqrt{s_{\mathrm{NN}}}$=5.02
  TeV}.
\newblock 5 2023, 2305.19060.

\bibitem{CMS:2023snh}
Armen Tumasyan et~al.
\newblock {Probing small Bjorken-$x$ nuclear gluonic structure via coherent
  J/$\psi$ photoproduction in ultraperipheral PbPb collisions at
  $\sqrt{s_\mathrm{NN}}$ = 5.02 TeV}.
\newblock 3 2023, 2303.16984.

\bibitem{Kowalski:2006hc}
H.~Kowalski, L.~Motyka, and G.~Watt.
\newblock {Exclusive diffractive processes at HERA within the dipole picture}.
\newblock {\em Phys. Rev. D}, 74:074016, 2006, hep-ph/0606272.

\bibitem{Marquet:2007nf}
Cyrille Marquet.
\newblock {A Unified description of diffractive deep inelastic scattering with
  saturation}.
\newblock {\em Phys. Rev. D}, 76:094017, 2007, 0706.2682.

\bibitem{Jones:2013pga}
S.P. Jones, A.D. Martin, M.G. Ryskin, and T.~Teubner.
\newblock {Probes of the small $x$ gluon via exclusive $J/\psi$ and $\Upsilon$
  production at HERA and the LHC}.
\newblock {\em JHEP}, 11:085, 2013, 1307.7099.

\bibitem{ALICE:2021tyx}
Shreyasi Acharya et~al.
\newblock {First measurement of the |$t$|-dependence of coherent $J/\psi$
  photonuclear production}.
\newblock {\em Phys. Lett. B}, 817:136280, 2021, 2101.04623.

\bibitem{Krelina:2018hmt}
Michal Krelina, Jan Nemchik, Roman Pasechnik, and Jan Cepila.
\newblock {Spin rotation effects in diffractive electroproduction of heavy
  quarkonia}.
\newblock {\em Eur. Phys. J. C}, 79(2):154, 2019, 1812.03001.

\bibitem{Li:2017mlw}
Yang Li, Pieter Maris, and James~P. Vary.
\newblock {Quarkonium as a relativistic bound state on the light front}.
\newblock {\em Phys. Rev. D}, 96:016022, 2017, 1704.06968.

\bibitem{Brodsky:1980vj}
Stanley~J. Brodsky, Tao Huang, and G.Peter Lepage.
\newblock {The Hadronic Wave Function in Quantum Chromodynamics}.
\newblock 6 1980.

\bibitem{Cox:2009ag}
B.E. Cox, J.R. Forshaw, and R.~Sandapen.
\newblock {Diffractive upsilon production at the LHC}.
\newblock {\em JHEP}, 06:034, 2009, 0905.0102.

\bibitem{Nemchik:1994fp}
J.~Nemchik, Nikolai~N. Nikolaev, and B.G. Zakharov.
\newblock {Scanning the BFKL pomeron in elastic production of vector mesons at
  HERA}.
\newblock {\em Phys. Lett. B}, 341:228--237, 1994, hep-ph/9405355.

\bibitem{Armesto:2014sma}
N\'estor Armesto and Amir~H. Rezaeian.
\newblock {Exclusive vector meson production at high energies and gluon
  saturation}.
\newblock {\em Phys. Rev. D}, 90(5):054003, 2014, 1402.4831.

\bibitem{Frankfurt:1996ri}
L.~Frankfurt, A.~Radyushkin, and M.~Strikman.
\newblock {Interaction of small size wave packet with hadron target}.
\newblock {\em Phys. Rev. D}, 55:98--104, 1997, hep-ph/9610274.

\bibitem{Goda:2022wsc}
Tomoki Goda, Krzysztof Kutak, and Sebastian Sapeta.
\newblock {Sudakov effects and the dipole amplitude}.
\newblock {\em Nucl. Phys. B}, 990:116155, 2023, 2210.16084.

\bibitem{Golec-Biernat:2017lfv}
Krzysztof Golec-Biernat and Sebastian Sapeta.
\newblock {Saturation model of DIS : an update}.
\newblock {\em JHEP}, 03:102, 2018, 1711.11360.

\bibitem{Azevedo:2022ozu}
Celsina~N. Azevedo, Victor~P. Goncalves, and Bruno~D. Moreira.
\newblock {Associated $\phi $ and $J/\Psi $ photoproduction in ultraperipheral
  PbPb collisions at the Large Hadron Collider and Future Circular Collider}.
\newblock {\em Eur. Phys. J. A}, 59(2):23, 2023, 2210.04861.

\bibitem{Mantysaari:2018nng}
Heikki M\"antysaari and Pia Zurita.
\newblock {In depth analysis of the combined HERA data in the dipole models
  with and without saturation}.
\newblock {\em Phys. Rev. D}, 98:036002, 2018, 1804.05311.

\bibitem{Chekanov:2004mw}
S.~Chekanov et~al.
\newblock {Exclusive electroproduction of J/psi mesons at HERA}.
\newblock {\em Nucl. Phys. B}, 695:3--37, 2004, hep-ex/0404008.

\bibitem{Alexa:2013xxa}
C.~Alexa et~al.
\newblock {Elastic and Proton-Dissociative Photoproduction of J/psi Mesons at
  HERA}.
\newblock {\em Eur. Phys. J. C}, 73(6):2466, 2013, 1304.5162.

\bibitem{Aktas:2005xu}
A.~Aktas et~al.
\newblock {Elastic J/psi production at HERA}.
\newblock {\em Eur. Phys. J. C}, 46:585--603, 2006, hep-ex/0510016.

\bibitem{TheALICE:2014dwa}
Betty~Bezverkhny Abelev et~al.
\newblock {Exclusive $\mathrm{J/}\psi$ photoproduction off protons in
  ultra-peripheral p-Pb collisions at $\sqrt{s_{\rm NN}}=5.02$ TeV}.
\newblock {\em Phys. Rev. Lett.}, 113(23):232504, 2014, 1406.7819.

\bibitem{Acharya:2018jua}
Shreyasi Acharya et~al.
\newblock {Energy dependence of exclusive $\mathrm {J}/\psi $ photoproduction
  off protons in ultra-peripheral p\textendash{}Pb collisions at
  $\sqrt{s_{\mathrm {\scriptscriptstyle NN}}} = 5.02$ TeV}.
\newblock {\em Eur. Phys. J. C}, 79(5):402, 2019, 1809.03235.

\bibitem{ALICE:2023mfc}
{Exclusive and dissociative J/$\psi$ photoproduction, and exclusive dimuon
  production, in p$-$Pb collisions at $\sqrt{s_{\rm NN}} = 8.16$ TeV}.
\newblock 4 2023, 2304.12403.

\bibitem{Aaij:2013jxj}
R~Aaij et~al.
\newblock {Exclusive $J/\psi$ and $\psi$(2S) production in pp collisions at $
  \sqrt{s} = 7$ TeV}.
\newblock {\em J. Phys. G}, 40:045001, 2013, 1301.7084.

\bibitem{Aaij:2018arx}
Roel Aaij et~al.
\newblock {Central exclusive production of $J/\psi$ and $\psi(2S)$ mesons in
  $pp$ collisions at $\sqrt{s}=13~$TeV}.
\newblock {\em JHEP}, 10:167, 2018, 1806.04079.

\bibitem{Schmidt:2001tu}
D.~Schmidt.
\newblock {\em {Diffractive photoproduction of charmonium in the H1 detector at
  HERA}}.
\newblock PhD thesis, Hamburg U., 2001.

\bibitem{Adloff:2002re}
C.~Adloff et~al.
\newblock {Diffractive photoproduction of psi(2S) mesons at HERA}.
\newblock {\em Phys. Lett. B}, 541:251--264, 2002, hep-ex/0205107.

\bibitem{Stasto:2000er}
A.~M. Stasto, Krzysztof~J. Golec-Biernat, and J.~Kwiecinski.
\newblock {Geometric scaling for the total gamma* p cross-section in the low x
  region}.
\newblock {\em Phys. Rev. Lett.}, 86:596--599, 2001, hep-ph/0007192.

\bibitem{Mueller:2018ned}
A.~H. Mueller and S.~Munier.
\newblock {Rapidity gap distribution in diffractive deep-inelastic scattering
  and parton genealogy}.
\newblock {\em Phys. Rev. D}, 98(3):034021, 2018, 1805.02847.

\bibitem{Mueller:2002zm}
A.~H. Mueller and D.~N. Triantafyllopoulos.
\newblock {The Energy dependence of the saturation momentum}.
\newblock {\em Nucl. Phys. B}, 640:331--350, 2002, hep-ph/0205167.

\bibitem{Munier:2003sj}
S.~Munier and Robert~B. Peschanski.
\newblock {Traveling wave fronts and the transition to saturation}.
\newblock {\em Phys. Rev. D}, 69:034008, 2004, hep-ph/0310357.

\bibitem{Munier:2003vc}
S.~Munier and Robert~B. Peschanski.
\newblock {Geometric scaling as traveling waves}.
\newblock {\em Phys. Rev. Lett.}, 91:232001, 2003, hep-ph/0309177.

\bibitem{Ryskin:1992ui}
M.~G. Ryskin.
\newblock {Diffractive J / psi electroproduction in LLA QCD}.
\newblock {\em Z. Phys. C}, 57:89--92, 1993.

\bibitem{Flett:2020duk}
C.A. Flett, A.D. Martin, M.G. Ryskin, and T.~Teubner.
\newblock {Very low $x$ gluon density determined by LHCb exclusive $J/\psi$
  data}.
\newblock 6 2020, 2006.13857.

\bibitem{Jones:2013eda}
S.P. Jones, A.D. Martin, M.G. Ryskin, and T.~Teubner.
\newblock {Predictions of exclusive \ensuremath{\psi}(2S) production at the
  LHC}.
\newblock {\em J. Phys. G}, 41:055009, 2014, 1312.6795.

\bibitem{Flett:2021ghh}
C.~A. Flett, J.~A. Gracey, S.~P. Jones, and T.~Teubner.
\newblock {Exclusive heavy vector meson electroproduction to NLO in collinear
  factorisation}.
\newblock {\em JHEP}, 08:150, 2021, 2105.07657.

\bibitem{Eskola:2022vpi}
Kari~J. Eskola, Christopher~A. Flett, Vadim Guzey, Topi L\"oyt\"ainen, and
  Hannu Paukkunen.
\newblock {Exclusive J/\ensuremath{\psi} photoproduction in ultraperipheral
  Pb+Pb collisions at the CERN Large Hadron Collider calculated at
  next-to-leading order perturbative QCD}.
\newblock {\em Phys. Rev. C}, 106(3):035202, 2022, 2203.11613.

\bibitem{Eskola:2022vaf}
Kari~J. Eskola, Christopher~A. Flett, Vadim Guzey, Topi L\"oyt\"ainen, and
  Hannu Paukkunen.
\newblock {Next-to-leading order perturbative QCD predictions for exclusive
  J/\ensuremath{\psi} photoproduction in oxygen-oxygen and lead-lead collisions
  at energies available at the CERN Large Hadron Collider}.
\newblock {\em Phys. Rev. C}, 107(4):044912, 2023, 2210.16048.

\bibitem{Lansberg:2021vie}
Jean-Philippe Lansberg, Maxim Nefedov, and Melih~A. Ozcelik.
\newblock {Matching next-to-leading-order and high-energy-resummed calculations
  of heavy-quarkonium-hadroproduction cross sections}.
\newblock {\em JHEP}, 05:083, 2022, 2112.06789.

\bibitem{Lansberg:2023kzf}
Jean-Philippe Lansberg, Maxim Nefedov, and Melih~A. Ozcelik.
\newblock {Curing the high-energy perturbative instability of
  vector-quarkonium-photoproduction cross sections at order $\alpha \alpha_s^3$
  with high-energy factorisation}.
\newblock 6 2023, 2306.02425.

\bibitem{ZEUS:2022sxn}
I.~Abt et~al.
\newblock {Measurement of the cross-section ratio
  \ensuremath{\sigma}$_{\Psi(2S)}$/\ensuremath{\sigma}$_{J/\Psi(1S)}$ in
  exclusive photoproduction at HERA}.
\newblock {\em JHEP}, 12:164, 2022, 2206.13343.

\bibitem{H1:2002yab}
C.~Adloff et~al.
\newblock {Diffractive photoproduction of psi(2S) mesons at HERA}.
\newblock {\em Phys. Lett. B}, 541:251--264, 2002, hep-ex/0205107.

\bibitem{Accardi:2012qut}
A.~Accardi et~al.
\newblock {Electron Ion Collider: The Next QCD Frontier}: {Understanding the
  glue that binds us all}.
\newblock {\em Eur. Phys. J. A}, 52(9):268, 2016, 1212.1701.

\bibitem{AbdulKhalek:2021gbh}
R.~Abdul~Khalek et~al.
\newblock {Science Requirements and Detector Concepts for the Electron-Ion
  Collider}: {EIC Yellow Report}.
\newblock {\em Nucl. Phys. A}, 1026:122447, 2022, 2103.05419.

\bibitem{AbdulKhalek:2022hcn}
R.~Abdul~Khalek et~al.
\newblock {Snowmass 2021 White Paper: Electron Ion Collider for High Energy
  Physics}.
\newblock 3 2022, 2203.13199.

\end{thebibliography}

\end{document}